\documentclass[12pt]{article}
\usepackage{amsfonts}
\usepackage{amssymb}
\usepackage{amsthm} 
\usepackage{newlfont}
\usepackage{epsfig}
\usepackage{amscd}
\usepackage{color}

\newcommand{\beq}{\begin{equation}}
\newcommand{\eeq}{\end{equation}}
\newcommand{\ba}{\begin{array}}
\newcommand{\ea}{\end{array}}
\newcommand{\bea}{\begin{eqnarray}}
\newcommand{\eea}{\end{eqnarray}}

\begin{document}
\begin{center}
{\large \sc \bf On  integration of a multidimensional version of  $n$-wave type equation}

\vskip 15pt

{\large A. I. Zenchuk
}

\vskip 8pt

{\it 
Institute of Problems of Chemical Physics, RAS,
Acad. Semenov av., 1
Chernogolovka,
Moscow region
142432,
Russia}

\smallskip

\vskip 5pt

e-mail:  {\tt zenchuk@itp.ac.ru
}

\vskip 5pt

{\today}

\end{center}

\begin{abstract}
We represent a version of multidimensional quasilinear partial differential equation (PDE) 
together with large manifold of particular solutions given in an integral form.  
The dimensionality of constructed PDE 
can be arbitrary. 
We call it the $n$-wave type PDE, although the structure of its nonlinearity   differs from that of the
classical completely integrable  (2+1)-dimensional $n$-wave equation. The richness of   solution space to such a PDE 
is characterized by a set  of arbitrary 
functions of several variables. However, this richness is not enough to provide the complete integrability, 
which is shown explicitly. We describe a 
class of multi-solitary wave  solutions in details.  Among examples of explicit particular solutions, we 
represent a lump-lattice solution depending 
on five independent 
variables. In Appendix, as an important  supplemental material, we 
show that our nonlinear PDE is reducible from the more general multidimensional PDE which can be derived using 
the dressing method based on the linear integral equation with the kernel of a special type
(a modification of the $\bar\partial$-problem).
The dressing algorithm gives us a key for construction of  higher order PDEs, 
although they are not discussed in this paper. 
\end{abstract}

\section{Introduction}
\label{Section:introduction}
A relevance of inverse spectral transform method (ISTM) \cite{GGKM,ZMNP,AC} 
is provided by the wide physical applicability of (1+1)- and (2+1)-dimensional nonlinear
ISTM-integrable 
partial differential equations   (PDEs), or soliton equations. 
As a well known realization  of  ISTM, we recall the 
dressing method having several versions \cite{ZS1,ZS2,ZM,BM,K}. 
However, the higher dimensional analogies of the classical  (1+1)- and (2+1)-dimensional 
soliton equations may not be constructed in a simple way. There are some examples of such 
multidimensional systems (see, for instance, \cite{AC}), but their solution spaces are 
restricted to (2+1)-dimensional manifolds. A possible 
way of increasing the dimensionality of solvable PDEs 
(together with the dimensionality of their solution space) was proposed in  \cite{ZS},
where a new class of partially integrable multidimensional 
PDEs was derived using the dressing method with the degenerate kernel of  integral operator.
The nonlinear PDEs from this class are well-structured, although their applicability requires additional
studies.

Regarding the non-soliton types of multidimensional integrable systems, we notice
the first order PDEs integrable by the method of characteristics 
(equations with wave breaking solutions) \cite{Whitham}, their matrix generalization \cite{SZ}, 
systems of hydrodynamic type \cite{T1,DN,T2,F},
 linearizable ($C$-integrable) models \cite{HopfCole,Calogero1,Calogero2,Calogero3,Calogero4}, 
linearizable equations with constraints \cite{Santini}, self-dual type 
equations (instanton equations) \cite{W,BZ}, nonlinear equations associated with 
commuting vector fields \cite{MS1,MS2,KAR}.
 Combinations  of different integration algorithms were used for construction of other types of 
 multidimensional PDEs in \cite{Z3,Z4}.

 This paper is devoted to a  multidimensional generalization of (2+1)-dimensional 
  $n$-wave type equation,
\cite{Kaup1,Kaup2}
\begin{eqnarray}\label{Nwave}
 [C^{(3)},U_{t_2}]-[C^{(2)},U_{t_3}] + C^{(2)} U_{t_1} C^{(3)} -
 C^{(3)} U_{t_1} C^{(2)}+ [[C^{(3)},U],[C^{(2)},U]] =0,
 \end{eqnarray}
which  is a well known example of ISTM-integrable  matrix $N\times N$ systems. 
Up to now, its  multidimensional  integrable 
generalization  is not found. Therewith, the 
physical applications of such a generalization are evident in the  areas where the  ($2+1$)-dimensional $n$-wave 
equation is  acknowledge. In particular, this equation appears in multiple-scale expansions of known physical 
systems. 
 
 Hereafter we will use the term 
 ''dimensionality'' in three different meanings. The first one is the number 
 of variables appearing in the nonlinear PDE (we refer to it as the $t$-dimensionality). 
 The second meaning is the matrix dimensionality of nonlinear PDE (the $N$-dimensionality). The third meaning is the 
 dimensionality of the solution space (i.e., the number of independent variables $t_i$
 which can be arbitrarily introduced in the solution space). It is referred to as the  $s$-dimensionality. 

 To analyze  the similarity and principal difference between the new PDE and the classical 
(2+1)-dimensional  $n$-wave equation,
let us   consider the linearized versions of  both of them, which read, respectively, 
\begin{eqnarray}\label{lineq}
 [C^{(3)},U_{t_2}]-[C^{(2)},U_{t_3}] + C^{(2)} U_{t_1} C^{(3)} -
 C^{(3)} U_{t_1} C^{(2)}=0,
\end{eqnarray}
and 
\begin{eqnarray}\label{Lin_Lim}
\sum_{i=1}^{\tilde D} \hat L^{(i)} U_{t_i} \hat R^{(i)}=0.
\end{eqnarray}
Both of these equations are matrix $N\times N$ PDEs.
The $t$-dimensionalities of these equations are, respectively, 3 and $\tilde D$, therewith 
the $s$-dimensionality of each of them  
is  less than the appropriate  $t$-dimensionality by one:
\begin{eqnarray}\label{tsdim}
{\mbox{$s$-dim. $=$ $t$-dim. }}-1,
\end{eqnarray}
which indicates the fullness of the solution spaces of the linear equations (\ref{lineq}) and (\ref{Lin_Lim}).
We emphasize, that the $N$-dimensionality is independent on the both $t$- and $s$-dimensionalities
and can be arbitrary.

It is remarkable, that the nonlinearity can be introduced into eq.(\ref{lineq}) in an integrable way
(resulting in eq.(\ref{Nwave})), i.e.,
preserving the above mutual relations among three dimensionalities.
The same has to be valid for the  desirable integrable 
non-linearization of eq.(\ref{Lin_Lim}).

 However, in this paper, the algorithm introducing the nonlinear terms in eq.(\ref{Lin_Lim}) 
 destroys the above  relations among three dimensionalities. 
In the new nonlinear PDE, both the $t$- and $N$-dimensionalities  increase with an increase in the $s$-dimensionality:
 $t$-dim. $\sim$ $(s$-dim.$)^2$, $N$-dim. $\sim$ $s$-dim.,
 so that $N$-dim. $\sim$ ($t$-dim.$)^2$.
We see that there are  mutual relations  among  all three dimensionalities.
Therewith,
the $t$- and $N$- dimensionalities  increase very fast with an increase in the $s$-dimensionality.  
As a consequence,
 the complete integrability of such a PDE may not be achieved.
 Thus, we deal with the partial integrability, 
 similar to ref.\cite{ZS}. 
 In addition, the large $t$-dimensionality requires the large $N$-dimensionality 
 and is not suitable for small $N$.
 Consequently, a valuable problem is compressing of the increase in the $t$- and/or $N$-dimensionalities 
 with an increase in the 
 $s$-dimensionality.
In addition the derived nonlinear PDE has the diagonal part, 
 unlike classical equation (\ref{Nwave}).
 
 Of course, there might be another algorithm introducing the nonlinearity in eq.(\ref{Lin_Lim}) 
 in a way preserving 
 the complete integrability of the resulting nonlinear PDE, but such an algorithm is not found up to now.
 
In this paper, we first write the nonlinear PDE itself and explicitly describe its available 
solution space  in terms of  the integral operator
with the  kernel of a special type. This representation of solution involves  arbitrary functions of 
 several independent variables mentioned above. 
 In principle, the  integral representation of solution  can 
be considered as a prescribed representation of the available solution space. But it is important to 
emphasize that 
both PDE and formula for its solution were derived from the more general formulas which, in turn, 
can be derived  via the dressing technique based on  the 
linear integral equation of  special form (a modification of the $\bar\partial$-problem \cite{ZM,BM,K}). 
The derivation algorithm is straightforward, but rather long, so 
that we do not fix the reader's attention on it and represent this algorithm  in Appendix. 
The meaning of  this algorithm is that it not only 
specifies the origin of  the proposed formulas for the  first order  
multidimensional quasilinear PDE together with
its solution manifold 
but also gives a key for construction of higher order PDEs,
which remain beyond the scope of this paper.

The structure of this paper is following. In Sec.\ref{Section:results}, we
represent the general form of nonlinear PDE derived in this paper together with the general 
formula for available solution manifold. 
 In Sec.\ref{Section:solutions}, we discuss the available  
manifold of explicit solutions for the nonlinear PDE (in particular, multi-solitary wave solutions) 
and construct some explicit solutions, for instance, the lump-lattice solution with five independent 
variables for 10-dimensional PDE.
 Results are discussed in Sec.\ref{Section:conclusion}. 
A dressing algorithm used for   derivation of the discussed nonlinear PDE is given 
in Appendix A, Sec.\ref{Section:appendix}. The richness of the solutions space is discussed in
Appendix B, Sec.\ref{Section:appendixB}. Relations among the $t$-, $N$- and $s$-dimensionalities 
are described in Appendix C, Sec.\ref{Section:appendixC}. 
The link to  
the classical completely integrable (2+1)-dimensional $n$-wave 
equation  is given in Appendix D (Sec.\ref{Section:appendixD}). 
Brief remarks on construction of higher order multidimensional nonlinear PDEs are given in Appendix E, 
Sec.\ref{Section:appendixE}.

\section{Partially integrable quasilinear matrix first order PDE in arbitrary dimensions}
\label{Section:results}
\label{Section:reduction}

The nonlinear PDE represented below is a particular realization of the 
multidimensional generalization of 
 $n$-wave   equation. The structure of its nonlinear term differs 
from that used in the classical case, but this equation is likely to play an
important  role in the multiple-scale expansions of real physical systems, like 
the (2+1)-dimensional $n$-wave equation. The study of 
multiple-scale expansion remains beyond the scope of this paper.

We propose the following  version of the $D^2$-dimensional matrix  $N\times N$ 
nonlinear PDE:
\begin{eqnarray}
\label{S_Q_simple_d_nl_U_h_matr}
\sum_{m_1,m_2=1}^{D} \hat L^{(m_1)}\left(  U_{t_{m_1m_2}} + 
U p \,p^T \hat a^{(1;m_11 )}\hat a^{(1;1m_2 )} s^{-1} U
- Us^{-1} \hat a^{(1;m_11 )}\hat a^{(1;1m_2 )} p \,p^T
 U 
\right)\hat R^{(m_2)} =0,
\end{eqnarray}
where $D$ and $N$ are some integer parameters,
 $p=(\underbrace{1\;\dots\;1}_N)^T$ is $N\times 1$ matrix,  $\hat L^{(m_1)}$, $\hat R^{(m_2)}$ 
 and $s$ are arbitrary constant diagonal matrices. The diagonal matrices $\hat a^{(1;m_1 1 )}$
 and $\hat a^{(1;1m_2 )}$ are related with $\hat L^{(m_1)}$ and  $\hat R^{(m_2)}$ 
 via the following system
 of linear algebraic equations:
\begin{eqnarray}\label{Z2gen_red_comp2}
&&
\sum_{m_1=1}^{D}\hat L^{(m_1)}_{\alpha\pm  i}
 \hat  a^{(1;m_11)}_{\alpha} =0, \\\label{tZ2gen_red_comp2}
 &&
\sum_{m_1=1}^{D}
 \hat  a^{(1;1m_1)}_{\alpha}\hat R^{(m_1)}_{\alpha\pm  i} =0,\;\;i=0,1,\dots,K-1,\\\label{DR}\label{Dmin}
 && D= 2 K,\;\; \hat L^{(1)}=\hat R^{(1)} =I_N,
\end{eqnarray}
where $I_N$ is the $N\times N $  identity matrix and, for any $N\times N$ diagonal  matrix $b$, we assume
\begin{eqnarray}
 &&
b^{(1;m_11)}_{N+i} =b^{(1;m_11)}_i,\;\;
 b^{(1;m_11)}_{-i} =b^{(1;m_11)}_{N-i},\;\;
i=0,1,\dots, K-1.
\end{eqnarray}

\subsection{Integral representation of solution}

For the sake of brevity,  we denote the integration over 
the  space of $K$-dimensional complex  spectral parameter $\mu$,
\begin{eqnarray}\label{spectral_par}
\mu=\{\mu_1,\dots,\mu_K\},
\end{eqnarray}
by $*$, i.e., for any  two functions $f(\mu)$ and $g(\mu)$ we have
\begin{eqnarray}
\label{def_ast}
f(\mu)*g(\mu)=\int f(\mu)g(\mu) d\Omega(\mu),
\end{eqnarray}
where  $\Omega(\mu)$ is some measure in the $K$-dimensional space of  parameter $\mu$. We also introduce the 
unit  ${\cal{I}}(\lambda,\mu)$ 
operator in a usual way:
\begin{eqnarray}\label{def_unit}
&&
f(\lambda,\nu)*{\cal{I}}(\nu,\mu) ={\cal{I}}(\lambda,\nu)*f(\nu,\mu)=f(\lambda,\mu).
\end{eqnarray}
If $d\Omega(\mu)=d\mu d\bar\mu$ and the integration is over the whole space of the complex vector parameter $\mu$, then
${\cal{I}}(\lambda,\mu)=\delta(\lambda-\mu)=\prod_{i=1}^K \delta(\lambda_i-\mu_i)$. 
We use the double index for the   independent variables $t_{m_1m_2}$
of the nonlinear PDE and denote the whole set of these variables  by  $t$. 

By direct substitution, we may verify that  eq.(\ref{S_Q_simple_d_nl_U_h_matr}) possesses a 
 solution $U(t)$ given in the following integral form
\begin{eqnarray}
\label{tVU}
U(t)=\frac{s \tilde V(t)s}{1+Q} ,\;\;Q=p^T s^2 p,
\end{eqnarray}
where
\begin{eqnarray}\label{tV0}
\label{tV}
&&
\tilde V(t)=\\\nonumber
&&
-2\sum_{k=1}^K (\Pi^T)^{k-1} g^{(k)}(\lambda)*R(\lambda,\mu)*
\Big(\Psi_0(\mu,\tilde\mu;t)*R(\tilde\mu,\nu)+{\cal{I}}(\mu,\nu)\Big)^{-1}*
\sum_{i=1}^K  g^{(i)}(\nu)\Pi^{i-1},
\end{eqnarray}
\begin{eqnarray}\label{RRR}
R(\lambda,\mu) ={\cal{I}}(\lambda,\mu) +\sum_{i,j=1}^K g^{(i)}(\lambda) \Pi^{i-1} s p p^T s (\Pi^T)^{j-1} 
g^{(j)}(\mu),
\end{eqnarray}
\begin{eqnarray}\label{Pi}
\Pi=\left(\begin{array}{ccccc}
0&1&0&0&\cdots\cr
0&0&1&0&\cdots\cr
0&0&0&1&\cdots\cr
\cdots&\cdots&\cdots&\cdots&\cdots\cr
1&0&0&0&\cdots
\end{array}\right),\;\;\Pi^0 =I_N.
\end{eqnarray} 
Therewith, the functions $g^{(i)}(\lambda)$ are arbitrary diagonal functions of spectral parameters satisfying the
 conditions
\begin{eqnarray}\label{varkapij}
|g^{(i)}_\alpha*g^{(j)}_\alpha|<\infty,\;\;i,j=1,\dots,K,\;\;\alpha =1,\dots,N,
\end{eqnarray}
and the function $\Psi_0$ is defined as follows:
 \begin{eqnarray}
 \label{G_Psi00}
&&
\Psi_0(\lambda,\mu;t) = \varepsilon(\lambda;t) {\cal{C}}(\lambda,\mu)  \varepsilon^{-1}(\mu;t).\\\label{G_e}
&&\varepsilon(\nu;t)=\exp
\left({\displaystyle\sum_{m_1,m_2=1}^D T^{(m_1m_2)}(\nu) t_{m_1m_2}}\right),
\end{eqnarray}
  where  $T^{(m_1m_2)}$ has the form
\begin{eqnarray}\label{T_red}
T^{(m_1m_2)}(\lambda) =\sum_{j=1}^K \tilde g^{(j)}(\lambda) 
\Pi^{j-1} \hat a^{(1;m_11)}  \hat a^{(1;1m_2)} (\Pi^T)^{j-1}
\end{eqnarray}
with
\begin{eqnarray}\label{tgr}
&&
\tilde g^{(1)}(\lambda) = 2\Big(
s+\sum_{i=2}^K \hat g^{(i)}(\lambda) \Pi^{i-1} s 
\Big)^{-1},\\\label{tgr1}
&&
\tilde g^{(j)}(\lambda) = \hat g^{(j)}(\lambda) \tilde g^{(1)}(\lambda)
,\;\;j=2,\dots,K,\\\label{hatgg}
&&
\hat g^{(i)}_\alpha (\lambda) = 
\frac{g^{(i)}_\alpha(\lambda)}{g^{(1)}_\alpha(\lambda)}, \;\;i=2,\dots,K, \;\;\hat g^{(1)}=1.
\end{eqnarray}
Formulas (\ref{T_red} -- \ref{hatgg}) mean that, in $T^{(m_1m_2)}$, there are 
   $(K-1)$ 
independent arbitrary  functions $\hat g^{(i)}(\lambda)$, $i=2,\dots,K$, of spectral parameters  
which characterize  the richness of the
solution space. Owing to them, the 
arbitrary functions of $2(K-1)$ independent variables $t_{m_1m_2}$ may be introduced into the solution space 
(see Sec.\ref{Section:appendixB} in Appendix for details).

\subsection{Hermitian reductions  $U= U^+$.}
\label{Section:Herm}
Eq.(\ref{S_Q_simple_d_nl_U_h_matr}) admits the Hermitian reduction 
\begin{eqnarray}\label{VH}
&&
U= U^+,\\
\label{LR_H}
&&
\hat R^{(m_1)}=\hat L^{(m_1)},\;\;{\mbox{Im}}\,L^{(m_1)}=0, \;\;t_{m_1m_2}=t_{m_2m_1} ,\\
\label{LR_H2}
&&
s=i s_0.
 \\\label{aa_h}
&&
\hat a^{(i;m_1 1)} = \hat a^{(i;1 m_1)},\;\;m_1=1,\dots, D,
\\\label{LR_H3}
&& \hat a^{(1,1m_1)} , \;s_0 \;\;\;{\mbox{are the real diagonal matrices}}.
\end{eqnarray}
Regarding the solution manifold, we have to add two more relations:
\begin{eqnarray}
\label{V02}
&&
{\cal{C}}^+(\bar\mu,\bar\lambda) ={\cal{C}}(\lambda,\mu),\\\label{gpl}
&&
\overline{g^{(k)}}(\bar \lambda) = g^k(\lambda).
\end{eqnarray}
The Hermitian  reduction  reduces the $t$-dimensionality from $D^2$ to 
$\frac{D(D+1)}{2}$.
Nonlinear PDE (\ref{S_Q_simple_d_nl_U_h_matr}) under this reduction  reads:
\begin{eqnarray}
\label{S_Q_simple_d_nl_U_h}
\sum_{m_1,m_2=1}^{D} \hat L^{(m_1)}\left(  U_{t_{m_1m_2}} + 
i U p \,p^T \hat a^{(1;m_11 )}\hat a^{(1;m_21 )} s^{-1}_0 U
- i Us^{-1}_0 \hat a^{(1;m_11 )}\hat a^{(1;m_21 )} p \,p^T
 U 
\right)\hat L^{(m_2)} =0.
\end{eqnarray}

\subsection{Example of eq.(\ref{S_Q_simple_d_nl_U_h}) with $K=2$
}
\label{Section:exampleK2}
In this subsection, we give  a brief analysis of PDE (\ref{S_Q_simple_d_nl_U_h}) associated with $K=2$ and 
different values of the $N$-dimensionality. Since $K=2$, then  $D=4$ owing to relation (\ref{Dmin}). 
The $t$-dimensionality is  
$D(D+1)/2=10$. In accordance
with eqs.(\ref{T_red}) and (\ref{tgr}), there is one arbitrary function of
spectral parameters in $T^{(m_1m_2)}$, and, consequently,
arbitrary functions of two independent variables may be introduced in solution space (the $s$-dim.$ =2$).  
If $N \le 3$, then the system is completely linear, which may be simply verified 
(the nonlinear terms disappear in virtue of 
eqs.(\ref{Z2gen_red_comp2},\ref{tZ2gen_red_comp2})). 
If $3< N<7$, then  matrix equation (\ref{S_Q_simple_d_nl_U_h}) can be splitted 
into two families of scalar PDEs. 
The first family consists of the  linear scalar PDEs for the set of scalar fields ${\cal{U}}^{(lin)}$. 
The second family consists of the non-linear scalar PDEs for another  set of scalar fields ${\cal{U}}^{(nl)}$.  
If the matrix dimensionality $N=4$, then only the diagonal part of eq.(\ref{S_Q_simple_d_nl_U_h}) is nonlinear, 
i.e., $\{U_{\alpha\alpha},\;\alpha=1,\dots,N\} \in {\cal{U}}^{(nl)} $,
$\{U_{\alpha\beta},\;\alpha\neq \beta,\;\alpha,\beta=1,\dots,N\} \in {\cal{U}}^{(lin)} $. 
If $N=5$, then the nonlinearity appears in the non-diagonal part, 
however this nonlinearity is trivial because each quadratic term of this nonlinearity  
involves fields from the set ${\cal{U}}^{(lin)}$ and consequently the sub-system of 
PDEs for the non-diagonal elements $U_{\alpha\beta}$ is essentially linear (the "hidden linearity").
The first nontrivial case corresponds to $N=6$. 
The nonlinear system consists of 21 independent 
scalar equations for the fields $U_{\alpha\beta}$, $\beta\ge \alpha$. 
Three equations are linear PDEs  and  18 are nonlinear PDEs, i.e
\begin{eqnarray} 
&&
\{U_{14},U_{25},U_{36} \}\in {\cal{U}}^{(lin)} ,\\\nonumber
&&
\{U_{\alpha\alpha},\;\;\alpha=1,\dots,6, \;\; U_{12}, U_{13}, U_{15}, U_{16},
U_{23},U_{24}, U_{26},U_{34}, U_{35}, U_{45},U_{46},U_{56} \}\in {\cal{U}}^{(nl)},
\end{eqnarray}
where the diagonal elements $U_{\alpha\alpha}$, $\alpha=1,\dots,6$, 
are the real fields and others are the complex ones.
Nonlinear system  (\ref{S_Q_simple_d_nl_U_h})  can be written as
\begin{eqnarray}\label{S_Q_simple_d_nl_U2}
&&
\sum_{m_1,m_2=1}^{4} \hat L^{(m_1)}_\alpha\left( ( U_{\alpha\beta})_{t_{m_1m_2}} - i
\sum_{{{\gamma,\delta=1}\atop{\delta\notin Z(\alpha,\beta,K)}}\atop{\gamma\neq \delta}}^6\frac{
\hat a^{(1;1m_2)}_\delta\hat a^{(1;m_11)}_\delta}{(s_0)_\delta}   \Big(U_{\alpha\gamma}U_{\delta\beta}
-
 U_{\alpha\delta}
  U_{\gamma\beta} \Big)
\right)\hat L^{(m_2)}_\beta =0,\;\;\;
\\\nonumber
&&
\alpha,\beta=1,\dots,6,\;\;\beta\neq \alpha+3,\\\label{S_Q_lin}
&&
\sum_{m_1,m_2=1}^{4} \hat L^{(m_1)}_\alpha ( U_{\alpha(\alpha+3)})_{t_{m_1m_2}}
\hat L^{(m_2)}_{\alpha+3},\;\;\alpha=1,2,3,
\end{eqnarray} 
where we introduce the set of indices $Z(\alpha,\beta,K)$:
\begin{eqnarray}\label{Zlist}
Z(\alpha,\beta,K) = \{\alpha\pm i, \beta \pm i, i=1,\dots,K-1\}.
\end{eqnarray}
If $U_{\alpha(\alpha+3)}=0$, $\alpha=1,2,3$, then  linear equations (\ref{S_Q_lin}) become identities and
one has the system of 18 scalar nonlinear  equations (\ref{S_Q_simple_d_nl_U2}).

\section{Construction of explicit solutions to eqs.(\ref{S_Q_simple_d_nl_U_h_matr}) and 
(\ref{S_Q_simple_d_nl_U_h})}
\label{Section:solutions}
In general, the integral operator $* (\Psi_0*R+{\cal{I}})$ in eqs.(\ref{tVU},\ref{tV}) can be inverted numerically. 
However, there is a particular case of degenerate kernel $\Psi_0$, when this operator may be 
inverted analytically resulting in the 
explicit formula for solution. 
Inverting the above integral operator is equivalent to solving the linear equation
\begin{eqnarray}\label{Psi_r}
{\cal{I}}(\lambda,\nu)=\theta(\lambda,\mu) * (\Psi_0(\mu,\tilde\mu)*R(\tilde\mu,\nu) + {\cal{I}}(\mu,\nu)) 
\end{eqnarray}
for the function $\theta$. The degenerate kernel  $\Psi_0(\lambda,\mu)$ means 
the degenerate matrix function 
 ${\cal{C}}(\nu,\mu)$:
\begin{eqnarray}
{\cal{C}}(\nu,\mu) =\sum_{i  =1}^{N_0} u^{(i)}(\nu) v^{(i)}(\mu),
\end{eqnarray}
where $N_0$ is some integer.
Then we may write
\begin{eqnarray}\label{Psi0sol}
&&
\Psi_0(\nu,\mu;t)*R(\mu,\lambda) =\sum_{i  =1}^{N_0} \phi^{(i)}(\nu;t) \psi^{(i)}(\lambda;t),\\\label{psiphi}
&&
 \phi^{(i)}(\nu;t) =\varepsilon(\nu;t)  u^{(i)}(\nu),\;\; \psi^{(i)}(\lambda) = 
\Big(v^{(i)}(\mu;t)  \varepsilon^{-1}(\mu;t)\Big) * R(\mu,\lambda).
\end{eqnarray}
Substituting eq.(\ref{Psi0sol}) into  eq.(\ref{Psi_r}) we obtain:
 \begin{eqnarray}\label{Psi_r2}
&&
{\cal{I}}(\lambda,\nu) = \theta(\lambda,\nu;t) +\sum_{i=1}^{N_0} \theta^{(i)}(\lambda,t)\psi^{(i)}(\nu;t),
\end{eqnarray}
where
\begin{eqnarray}
\theta^{(i)}(\lambda,t)= \theta(\lambda,\nu;t)*\phi^{(i)}(\nu;t).
\end{eqnarray}
The functions $\theta^{(i)}$ satisfy a system of  linear algebraic equations which can be derived
applying the operators $*\phi^{(k)}(\lambda;t)$ 
to  eq.(\ref{Psi_r2}). As a result, we obtain:
 \begin{eqnarray}\label{Psi_W}
&&
\phi^{(k)}(\lambda,t) = \theta^{(k)}(\lambda,t) +
\sum_{i=1}^{N_0} \theta^{(i)}(\lambda,t) S^{(ik)},\;\;k=1,\dots,N_0,
\end{eqnarray}
where
\begin{eqnarray}\label{sS}
S^{(ik)}(t)&=& \psi^{(i)}(\lambda;t)*\phi^{(k)}(\lambda;t) =\\\nonumber
&&
 v^{(i)}(\mu)*u^{(k)}(\mu)  +
v^{(i)}(\mu)*\Big( \varepsilon^{-1}(\mu;t) z(\mu)\Big) \Big(z^T(\lambda) 
\varepsilon(\lambda;t)\Big)*u^{(k)}(\lambda),
\end{eqnarray}
and we introduce the notation
\begin{eqnarray}\label{zz}
z(\lambda) = \sum_{k=1}^K g^{(k)}(\lambda) \Pi^{k-1} s
\end{eqnarray}
for the sake of brevity.
Eqs. (\ref{Psi_W}) represent the linear non-homogeneous system of $N_0$ matrix $N\times N$ 
equations for the functions $\theta^{(k)}$, $k=1,\dots N_0$. 
The solvability of this system  must be provided by the matrices $S^{(ik)}$. 
In the simplest case, $N_0=1$, eq.(\ref{Psi_W}) can be easily solved:
\begin{eqnarray}\label{N0_1}
\theta^{(1)}(\lambda,t)=\phi^{(1)}(\lambda,t) \Big(
I_N + S^{(11)}(t)
\Big)^{-1}.
\end{eqnarray}
After functions $\theta^{(k)}$ are found, eq.(\ref{Psi_r2}) yields us 
expression for $\theta$:
\begin{eqnarray}\label{Psi_W_expl}
 \theta(\lambda,\nu,t)={\cal{I}}(\lambda,\nu) -
 \sum_{i=1}^{N_0} \theta^{(i)}(\lambda,t) \psi^{(i)}(\nu;t).
\end{eqnarray}
 Finally, substituting $\theta$  from (\ref{Psi_W_expl}) 
 (with $\theta^{(i)}$ given as solutions to  system (\ref{Psi_W}) ) 
 instead of  the operator $*(\Psi*R+{\cal{I}})^{-1}$ in 
 eq.(\ref{tV}) and using expression (\ref{RRR}) for $R$,  we obtain:
\begin{eqnarray}\label{tV_f}
&&
\tilde V(t)=
-2\sum_{k=1}^K (\Pi^{k-1})^T g^{(k)}(\lambda) * \Big(
{\cal{I}}(\lambda,\nu) + r(\lambda)  z^T(\nu)\Big)* 
\sum_{j=1}^K g^{(j)}(\nu)\Pi^{j-1}    +\\\nonumber
&&
2\sum_{k=1}^K (\Pi^{k-1})^T g^{(k)}(\lambda) *\Big({\cal{I}}(\lambda,\mu) + 
z(\lambda)z^T(\mu)\Big)*\\\nonumber
&&
\sum_{i=1}^{N_0} \theta^{(i)}(\mu)\Big( v^{(i)}(\nu) \varepsilon^{-1}(\nu)\Big)*\Big({\cal{I}}(\nu,\tilde\nu)
+ z(\nu)  z^T(\tilde\nu)\Big) * \sum_{j=1}^K g^{(j)}(\tilde\nu)\Pi^{j-1}  .
\end{eqnarray}
Hereafter we take
\begin{eqnarray}
d\Omega(\lambda)=\prod_{i=1}^K d \lambda_i.
\end{eqnarray}
Next,  conditions (\ref{varkapij}) must be satisfied by a proper choice of the functions  
$g^{(k)}(\lambda)$, for instance,
\begin{eqnarray}\label{ggcc}
&&
g^{(i)}_\alpha(\lambda) =c^{(i)}_\alpha \lambda_i
\exp\frac{1}{2}\Big( -\sum_{j=1}^K \Big(w_1^2(Re\;\lambda_j)^2 + w_2^2 (Im\;\lambda_j)^2 \Big)\Big).
\end{eqnarray}
Here, $w_j$, $j=1,2$,  and $c^{(i)}$, $i=1,\dots,K$, are the real parameters.
In this case, we obtain the following values for the parameters 
$\varkappa^{(ij)}_\alpha=g^{(i)}_\alpha*g^{(j)}_\alpha$:
\begin{eqnarray}
\varkappa^{(ii)}_\alpha \equiv \varkappa^{(i )}_\alpha  =
\frac{(w_2^2-w_1^2)\pi^K (c^{(i)}_\alpha)^2}{2 (w_1 w_2)^{K+2}}  <\infty
,\;\;
\varkappa^{(ij)}_\alpha=0, \;\;i\neq j,
\end{eqnarray}
which must be substituted in eq.(\ref{tV_f}) (these parameters appear in the scalars $g^{(j)}*z$ 
and $z^T*g^{(j)}$).
Now we show that choice (\ref{ggcc}) does not reduce 
the richness of the solution space to the nonlinear PDE and provides the  existence 
of $(K-1)$ independent combinations of  complex parameters in 
$T^{(m_1m_2)}(\lambda)$. 
In fact, substituting eqs.(\ref{ggcc}) into eqs.(\ref{tgr},\ref{tgr1})
 we obtain:
\begin{eqnarray}\label{htg2red}
&&
\tilde g^{(1)}_\alpha(\lambda) =
 2  \left(  s_\alpha +  \sum_{i=2}^K 
\frac{c^{(i)}_\alpha \lambda_i}{ c^{(1)}_\alpha\lambda_1}     s_{\alpha+i-1}\right)^{-1}
,\\\nonumber
&&
\tilde g^{(j)}_\alpha(\lambda) =
 \frac{c^{(j)}_\alpha \lambda_j}{  c^{(1)}_\alpha) \lambda_1 } \tilde g^{(1)}_\alpha(\lambda)
,\;\;j=2,\dots,K,
\end{eqnarray}
which must be used in $T^{(m_1m_2)}$ defined by eq.(\ref{T_red}).
We see that the parameter $\lambda_1$ appears only in the 
ratios $\lambda_j/\lambda_1$ in eqs. (\ref{htg2red}) and, 
consequently, in $T^{(m_1m_2)}(\lambda)$. Thus,  
there are $K-1$ independent complex parameters $\tilde \lambda_i =\lambda_i/\lambda_1$, $i=2,\dots,K$ 
in the function $\varepsilon(\lambda;t)$.

\subsection{Multi-solitary wave  solutions to eq.(\ref{S_Q_simple_d_nl_U_h_matr})}
In order to obtain the multi-solitary wave solution, we have to introduce the $\delta$-functions in $v^{(i)}$ and $u^{(i)}$:
\begin{eqnarray}\label{uv}
u^{(i)}(\lambda) = \delta(\lambda- p_i),\;\; v^{(i)}(\mu) = v^{(i)}_0 \delta(\mu-q_i),
\end{eqnarray}
where
we use the following notations
\begin{eqnarray}
\delta(\lambda- p_i) =\prod_{j=1}^{K} \delta(\lambda_j- p_{ji}),\;\;
\delta(\mu- q_i) =\prod_{j=1}^{K} \delta(\mu_j- q_{ji}) .
\end{eqnarray}
Here,
 $v^{(i)}_0$ are constant matrices, while $p_i =\{p_{i1},\dots,p_{iK}\}$ and 
 $q_i=\{q_{i1},\dots,q_{iK }\}$ are some constant multi-component complex parameters, $p_i\neq q_j$,
$i,j=1,\dots, N_0$.

\subsubsection{Transformation of eq.(\ref{Psi_W})}
{{}
Now, substituting $\phi^{(k)}$ from the first of eqs.(\ref{psiphi}) 
in 
eqs.(\ref{Psi_W}), we transform the later equation to the following form:}
\begin{eqnarray}\label{Psi_W_sol}
&&
\varepsilon(p_k,t)\delta(\lambda-p_k) = \theta^{(k)}(\lambda,t) +
\sum_{i=1}^{N_0} \theta^{(i)}(\lambda,t) S^{(ik)},
\end{eqnarray}
where $S^{(ik)}$ are given by eq.(\ref{sS}) after substitutions (\ref{uv}):
\begin{eqnarray}\label{sS_sol}
S^{(ik)}(t)&=& 
v^{(i)}_0*\Big( \varepsilon^{-1}(q_i;t) z(q_i)\Big) \Big(z^T(p_k) 
\varepsilon(p_k;t)\Big),
\end{eqnarray}
and we take into account that $v^{(i)}*u^{(j)}=0$, $i,j=1,\dots, N_0$, which immediately follows from 
(\ref{uv}).
We may collect the $t$-dependence in the rhs of equation (\ref{Psi_W_sol})
multiplying it by $\varepsilon^{-1}(p_k,t)$ from the right and introducing the function
\begin{eqnarray}
\hat \theta^{(i)}(\lambda;t)=\theta^{(i)}(\lambda;t) v^{(i)}_0\varepsilon^{-1}(q_i).
\end{eqnarray}
Then,  eq.(\ref{Psi_W_sol}) yields
\begin{eqnarray}\label{Psi_W_sol2}
&&
\delta(\lambda-p_k) = \sum_{i=1}^{N_0}\hat \theta^{(i)}(\lambda,t) \hat S^{(ik)} ,\;\;k=1,\dots,N_0,
\end{eqnarray}
where
\begin{eqnarray}
\label{th}
\hat S^{(ik)}=\varepsilon(q_i,t) (v^{(i)}_0)^{-1} \varepsilon^{-1}(p_i,t)\delta_{ik}
+z(q_i) z^T(p_k).
\end{eqnarray}

\subsubsection{Non-singularity conditions for functions $\hat \theta^{(i)}$}

In general, functions $\hat\theta^{(i)}$ constructed as solutions of linear  system (\ref{Psi_W_sol2}) 
have singularities in the $t$-space, 
which lead to   singularities in the solution $U(t)$ of the nonlinear PDE. To derive the non-singularity
conditions, we first rewrite  system (\ref{Psi_W_sol2}) in the following row-matrix form:
\begin{eqnarray}\label{matr}
[\delta(\lambda-p_1) \cdots  \delta(\lambda-p_{N_0})] = 
[\hat \theta^{(1)} \cdots\hat \theta^{(N_0)}]^T {\cal{S}}, 
\end{eqnarray}
where the matrix ${\cal{S}}$ has the following block structure:
\begin{eqnarray}
{\cal{S}} =\left(
\begin{array}{cccc}
\hat S^{(11)} & \hat S^{(12)} & \cdots & \hat S^{(1N_0)}\cr
\hat S^{(21)} & \hat S^{(22)} & \cdots & \hat S^{(2N_0)}\cr
\cdots &\cdots &\cdots &\cdots \cr
\hat S^{(N_01)} & \hat S^{(N_02)} & \cdots &\hat  S^{(N_0N_0)}\cr
\end{array}
\right).
\end{eqnarray}
In other words, the elements of ${\cal{S}}$ read
\begin{eqnarray}\label{calS}
{\cal{S}}_{N(i-1) +\alpha ,N(j-1)+\beta} = \hat S^{(ij)}_{\alpha,\beta},\;\;i,j=1,\dots, N_0.
\end{eqnarray}
Functions $\hat \theta^{(i)}(\lambda,t)$ have no singularities if  
\begin{eqnarray}\label{Sdet}
|\det{\cal{S}}|\neq 0,\;\;\forall \,t.
\end{eqnarray}
Using eq.(\ref{th}), we may write the following expressions for the elements of  matrix ${\cal{S}}$:
\begin{eqnarray}
{\cal{S}}_{ij}= \epsilon_i \hat v^{-1}_{ij} \tilde\epsilon_j(t) + \hat r^{(1)}_i \hat r^{(2)}_j, 
\;\;i,j=1,\dots,N_0N,
\end{eqnarray}
where the elements of the matrices $\epsilon$, $\tilde\epsilon$, $\hat v$ and $\hat r^{(i)}$, $i=1,2$, are defined 
as 
\begin{eqnarray}
&&
\hat v_{N(i-1) +\alpha ,N(j-1)+\beta}=(v^{(i)}_0)_{\alpha,\beta} \delta_{ik},\\\nonumber
&&
\epsilon_{N(i-1) +\alpha}(t) =\varepsilon_\alpha (q_i;t) ,\;\;
\tilde\epsilon_{N(k-1) +\beta}(t) =\varepsilon^{-1}_\beta (p_k;t),\\\nonumber
&&
\hat r^{(1)}_{N(i-1) +\alpha}=(z(q_i) )_\alpha,\;\;
\hat r^{(2)}_{N(k-1) +\beta}=( z^T(p_k))_\beta.
\end{eqnarray}
Let us derive the constraints for the elements of $v_0^{(i)}$
and $\xi$ which  guaranty the validity of non-singularity condition (\ref{Sdet}). 
The later  has the following explicit form:
\begin{eqnarray}\label{detneq0gen0}
\left|\prod_{k,l=1}^{N_0 N}
\frac{\epsilon_k(t) \tilde\epsilon_l(t)}{\det \,\hat v}\left(1+ \sum_{i,j=1}^{N_0N}
 \frac{\hat r^{(1)}_i \hat r^{(2)}_j \hat v_{ji}}{ \epsilon_i(t) \tilde\epsilon_j(t)}\right)\right|\equiv
 \left|\prod_{k,l=1}^{N_0 N}
\frac{\epsilon_k(t) \tilde\epsilon_l(t)}{\det \,\hat v}
\right| \left|
\left(1+ \sum_{i,j=1}^{N_0N}
 \frac{\hat r^{(1)}_i \hat r^{(2)}_j \hat v_{ji}}{ \epsilon_i(t) \tilde\epsilon_j(t)}\right)\right|\neq 0.
\end{eqnarray}
In eq.(\ref{detneq0gen0}), the $t$-dependence appears only in $\epsilon_i$ and  
$\tilde\epsilon_j$. 
The first factor $\left|\prod_{k,l=1}^{N_0 N}
\frac{\epsilon_k(t) \tilde\epsilon_l(t)}{\det \,\hat v}
\right|$ is  positive for all $t$ if $\det\,\hat v\neq 0$, which will be assumed hereafter.
Thus, condition (\ref{calS}) reduces to 
\begin{eqnarray}\label{detneq0gen}
 \left|
1+ \sum_{i,j=1}^{N_0N}
 \frac{\hat r^{(1)}_i \hat r^{(2)}_j \hat v_{ji}}{ \epsilon_i(t) \tilde\epsilon_j(t)}\right|\neq 0.
\end{eqnarray} 
The analysis of condition (\ref{detneq0gen}) depends on whether the products $\epsilon_i \tilde \epsilon_j$ depend on $t$. 
We consider the real matrix $v_0$. Then
condition   (\ref{detneq0gen}) may be simply satisfied if the arguments of the matrix exponents 
$\epsilon_i$ and $\tilde\epsilon_i$ are either real or imaginary.
First, we assume that these arguments are real and consider two cases.

\noindent
1. Let the combinations $\epsilon_i\tilde\epsilon_j$ depend on $t$ for all $i$ and $j$. 
Condition (\ref{detneq0gen}) is satisfied if the coefficients ahead of all exponents are positive, i.e.
\begin{eqnarray}\label{gen11}
\hat r^{(1)}_i \hat r^{(2)}_j  \hat v_{ji} >0 ,\;\;\forall \, i,j.
\end{eqnarray}

\noindent
2. Let 
$\epsilon_i\tilde\epsilon_i=1$, and  the combinations $\epsilon_i\tilde\epsilon_j$ depend on $t$ for $i\neq j$.
Then 
condition (\ref{detneq0gen})
reads
\begin{eqnarray}\label{detneq0gen1}
 \left|
1+\sum_{i=1}^{N_0N}
 \hat r^{(1)}_i \hat r^{(2)}_i \hat v_{ii}+ \sum_{{i,j=1}\atop{i\neq j}}^{N_0N}
 \frac{\hat r^{(1)}_i \hat r^{(2)}_j \hat v_{ji}}{ \epsilon_i(t) \tilde\epsilon_j(t)}\right|\neq 0.
\end{eqnarray}
which holds if 
\begin{eqnarray}\label{gen12}
{\mbox{sign}} \Big(\hat r^{(1)}_i \hat r^{(2)}_j  \hat v_{ji}\Big) =
{\mbox{sign}} \Big(1+\sum_{\tilde i=1}^{N_0N} 
\hat r^{(1)}_{\tilde i} \hat r^{(2)}_{\tilde i}  \hat v_{\tilde i\tilde i}\Big) ,\;\;
i\neq j.
\end{eqnarray}

\noindent
Second, we assume that the 
arguments of  the exponents $\epsilon_i$ and $\tilde\epsilon_i$ are imaginary and consider two 
following cases.

\noindent
1. Let the combinations $\epsilon_i\tilde\epsilon_j$ depend on $t$ for all $i$ and $j$. 
Then the  oscillating terms in condition (\ref{detneq0gen}) have the amplitudes $|r^{(1)}_i r^{(2)}_j  \hat v_{ji}|$. 
Consequently, condition (\ref{detneq0gen})
holds if 
\begin{eqnarray}\label{gen21}
1-\sum_{i,j=1}^{N_0N}
\hat r^{(1)}_i \hat r^{(2)}_j  \hat v_{ji}>0 .
\end{eqnarray}

\noindent
2. Let 
$\epsilon_i\tilde\epsilon_i=1$, and the combinations $\epsilon_i\tilde\epsilon_j$ depend on $t$ for $i\neq j$.
Then 
condition (\ref{detneq0gen}) transforms to (\ref{detneq0gen1}), which holds if
\begin{eqnarray}\label{gen22}
{\mbox{abs}}\Big(1+\sum_{i,j=1}^{N_0N} 
\hat r^{(1)}_{\tilde i}\hat  r^{(2)}_{\tilde i}  \hat v_{\tilde i\tilde i} \Big) -  \sum_{{i,j=1}\atop{i\neq j}}^{N_0N} {\mbox{abs}} \Big(\hat r^{(1)}_i \hat r^{(2)}_j \hat v_{ji}\Big)>0 .
\end{eqnarray}
The analysis of condition (\ref{detneq0gen}) in general case  may be done similarly.  But it is more 
cumbersome and  we do not represent it here.  

\paragraph{General form of functions $\hat \theta^{(i)}$ as solutions of eq.(\ref{Psi_W_sol2}).}

The general solution $\hat \theta^{(i)}(\lambda,t)$ of eq.(\ref{Psi_W_sol2}) can be represented as a 
linear combination of  delta-functions:
\begin{eqnarray}
\label{thata_delta}
\hat\theta^{(i)}(\lambda,t) =\sum_{j=1}^{N_0} w^{(ji)}(t) \delta(\lambda-p_j),
\end{eqnarray}
where $w^{(ji)}$ are solutions to the following linear system
\begin{eqnarray}\label{w}
 \sum_{i=1}^{N_0} w^{(ji)}\hat S^{(ik)}  = \delta_{jk},\;\;j,k=1,\dots,N_0.
\end{eqnarray}
Then the function 
$\tilde V$ given by  formula
(\ref{tV_f}) reads:
\begin{eqnarray}
\label{tV_f_sol}
&&
\tilde V(t)=
C_0 + 2\sum_{n,i=1}^{N_0}C_{1n} w^{(ni)}(t) \tilde C_{1i},
 \end{eqnarray}
 where the constant matrices $C_0$, $C_{1n}$, and  $\tilde C_{1n}$ are given by the following expressions:
\begin{eqnarray}\label{C0}
C_0&=&-2\sum_{k=1}^K (\Pi^T)^{k-1}\varkappa^{(k)} \Pi^{k-1}-\\\nonumber
&&
2 \sum_{j=1}^K \sum_{k=1}^K(\Pi^T)^{k-1}   \varkappa^{(k)}\Pi^{k-1}  s p\, p^Ts^T (\Pi^T)^{j-1}   \varkappa^{(j)} \Pi^{j-1}
\\\label{C11}
C_{1n}&=&
 \sum_{k=1}^K (\Pi^T)^{k-1}  
\Big(g^{(k)}(p_n)+\varkappa^{(k)}\Pi^{k-1} s p p^Ts^T \sum_{l=1}^K (\Pi^T)^{l-1} g^{(l)}(p_n)\Big),\\
\label{tC1}
\tilde C_{1n}&=&
 \sum_{k=1}^K  
\Big(g^{(k)}(q_n)+
 \sum_{l=1}^K g^{(l)}(q_n)(\Pi)^{l-1} s p p^Ts^T(\Pi^T)^{k-1}\varkappa^{(k)} \Big)
(\Pi)^{k-1} .
 \end{eqnarray}

\subsubsection{One-solitary wave  solution.}
Similar to the classical integrable PDEs, we refer to the solution corresponding to the
case $N_0=1$ as  the one-solitary wave solution.
Let   
$v^{(1)}_0=v_0$, and $p_1$, $q_1$ be real.
Then eq.(\ref{N0_1}) yields
\begin{eqnarray}\label{theta}
\theta^{(1)}(\lambda,t)= \delta(\lambda-p_1)\theta^{(1)}_0(t),\;\;\;\theta^{(1)}_0(t)=
\varepsilon(p_1;t)\Big(1+v_0\varepsilon^{-1}(q_1;t)z(q_1)  z^T(p) \varepsilon(p_1,t)\Big)^{-1},
\end{eqnarray}
so that the solution of eq.(\ref{w}) reads: 
\begin{eqnarray}\label{w11}
w^{(11)}=\theta^{(1)}_0 v_0 \varepsilon^{-1}(q_1).
\end{eqnarray}
Now, substituting this $w^{(11)}$ in eq.(\ref{tV_f_sol}), we obtain 
\begin{eqnarray}\label{tVv}
&&
\tilde V(t)=
C_0 +
2C_{11}
\hat\theta_0(t)
\tilde C_{11},
 \end{eqnarray}
where 
\begin{eqnarray}\label{thetav}
&&
\hat\theta_0(t) =(\hat S^{(11)})^{-1}= \Big(\varepsilon(q_1;t) v_0^{-1} \varepsilon^{-1}(p_1;t)+ 
z(q_1)  z^T(p_1)\Big)^{-1},
 \end{eqnarray}
 and the constant matrix $C_0$ is given by eq.(\ref{C0}), while expressions (\ref{C11})  for  $ C_{11} $ 
and  (\ref{tC1})  for
$\tilde C_{11} $ are following:
\begin{eqnarray}\label{tC1_onesol}
C_{11}&=&
\sum_{j=1}^K \sum_{k=1}^K (\Pi^T)^{k-1} 
\Big(g^{(k)}(p_1)+\varkappa^{(k)}\Pi^{k-1} s p\, p^Ts^T \sum_{l=1}^K (\Pi^T)^{l-1} g^{(l)}(p_1)\Big),\\
\nonumber
\tilde C_{11}&=&
 \sum_{k=1}^K  
\Big(g^{(k)}(q_1)+
 \sum_{l=1}^K g^{(l)}(q_1)(\Pi)^{l-1} s p p^Ts^T(\Pi^T)^{k-1}\varkappa^{(k)} \Big)
(\Pi)^{k-1} .
 \end{eqnarray}
 In eq.(\ref{thetav}), $\varepsilon$ is given by  eq.(\ref{G_e})
 with $T^{(m_1m_2)}$ from eq.(\ref{T_red}) and $\tilde g^{(i)}$ from eq.(\ref{htg2red}).
 
In the case $N_0=1$,  non-singularity condition (\ref{detneq0gen}) 
reads  as 
\begin{eqnarray}\label{detneq0}
\left|
1+\sum_{\alpha,\beta=1}^N \frac{ (z(q_1))_\alpha (z(p_1))_\beta  
(v_0)_{\beta\alpha}}{\varepsilon_\alpha(q_1;t)\varepsilon^{-1}_\beta(p_1;t)}
\right|
\neq 0\;\;\forall \, t.
\end{eqnarray}
We may simply satisfy this condition, if  the arguments of  matrix exponents   $\epsilon_\alpha(p_1;t)$ 
and $\epsilon_\beta(q_1;t)$ ($\alpha,\beta=1,\dots,N$)
are either real or imaginary, rewriting eqs.(\ref{gen11}-\ref{gen22}) for the case $N_0=1$ as follows.  

Assuming that  the arguments of the above matrix exponents
are real, we  obtain two following cases.

\noindent
1. Let $\varepsilon_\alpha(q_1;t)\varepsilon^{-1}_\beta(p_1;t)$ depend on $t$ for all $\alpha$ and $\beta$.
Then condition (\ref{gen11})
reads
\begin{eqnarray}\label{one11}
(z(q_1))_\alpha (z(p_1))_\beta  (v_0)_{\beta\alpha} >0 ,\;\;\forall \, \alpha,\beta.
\end{eqnarray}

\noindent
2. Let
$\varepsilon_\alpha(q_1;t)\varepsilon^{-1}_\alpha(p_1;t) =I_N$ and 
$\varepsilon_\alpha(q_1;t)\varepsilon^{-1}_\beta(p_1;t)$ depend on $t$ for all $\alpha\neq \beta$.
Then 
condition (\ref{gen12})
reads
\begin{eqnarray}\label{one12}
{\mbox{sign}} (z(q_1))_\alpha (z(p_1))_\beta  (v_0)_{\beta\alpha} =
{\mbox{sign}} (1+\sum_{\gamma=1}^N 
(z(q_1))_{\gamma} (z(p_1))_{\gamma}  (v_0)_{\gamma\gamma}) ,\;\;
\forall \,\alpha\neq \beta.
\end{eqnarray}
If the matrix exponents have the imaginary arguments, we  obtain
two other cases.

\noindent
1.  Let $\varepsilon_\alpha(q_1;t)\varepsilon^{-1}_\beta(p_1;t)$ depend on $t$ for all $\alpha$ and $\beta$.
Then condition (\ref{gen21})
reads
\begin{eqnarray}\label{one21}
1-\sum_{\alpha,\beta=1}^N
(z(q_1))_\alpha (z(p_1))_\beta  (v_0)_{\beta\alpha}>0.
\end{eqnarray}

\noindent
2. Let
$\varepsilon_\alpha(q_1;t)\varepsilon^{-1}_\alpha(p_1;t) =I_N$ and $\varepsilon_\alpha(q_1;t)
\varepsilon^{-1}_\beta(p_1;t)$ depend on $t$ for all $\alpha\neq \beta$.
Then 
condition (\ref{gen22})
reads
\begin{eqnarray}\label{one22}
{\mbox{abs}} \Big(1+\sum_{ \alpha=1}^N 
(z(q_1))_{\alpha} (z(p_1))_{\alpha}  (v_0)_{\alpha\alpha} \Big)
- \sum_{{\alpha,\beta=1}\atop{\alpha\neq\beta}}^N {\mbox{abs}} \Big( (z(q_1))_\alpha 
(z(p_1))_\beta  (v_0)_{\beta\alpha}\Big) >0.
\end{eqnarray}

\subsection{Hermitian reduction}
Hermitian reduction (\ref{VH}) requires relations ({\ref{LR_H}-\ref{gpl}), 
so that 
eqs.(\ref{htg2red}) for  $\tilde g^{(i)}$ read 
\begin{eqnarray}\label{htg3red}
&&
\tilde g^{(1}_\alpha(\lambda) =
 -2i \left( (s_0)_\alpha +  \sum_{i=2}^K 
\frac{c^{(i)}_\alpha \lambda_i}{c^{(1)}_\alpha \lambda_1}    (s_0)_{\alpha+i-1}\right)^{-1}
,\\\nonumber
&&
\;\;\tilde g^{(j)}_\alpha(\lambda) = \frac{c^{(j)}_\alpha\lambda_j}{c^{(1)}_\alpha\lambda_1 }
\tilde g^{(1}_\alpha(\lambda)
,\;\;j=2,\dots,K.
\end{eqnarray}

\subsection{Examples of explicit solutions}
Now we turn to   eq.(\ref{S_Q_simple_d_nl_U2}) and consider the solution  corresponding to
 $K=2$, $D=4$, $N=6$.  The $t$-dimensionality is 
 $D(D+1)/2=10$ in this case. We fix the parameters $L^{(m_1)}$ and $s_0$ as  follows:
\begin{eqnarray}
L^{(1)}=I_N,\;\;L^{(m_1)}_\alpha =\left\{
\begin{array}{ll}
\alpha^{m_1-1},& m_1=2,3,4,\;\;\alpha=1,2,3\;\;\cr
(2-\alpha)^{m_1-1},& m_1=2,3,4,\;\;\alpha=4,5,6\;\;\cr
\end{array}\right.,\;\;s_0=I_N.
\end{eqnarray}
 Elements of the matrices $\hat a^{(1;1m_1)}\equiv \hat a^{(1;m_11)}$, $m_1=2,3,4$, 
may be found from eqs.(\ref{Z2gen_red_comp2}) and (\ref{tZ2gen_red_comp2}):
\begin{eqnarray}
&&
\hat a^{(1;12)}={\mbox{diag}}(-\frac{5}{4},-\frac{11}{6} ,-\frac{1}{3} ,\frac{1}{2},
\frac{13}{12},-\frac{5}{12}),\;\;
\hat a^{(1;13)}={\mbox{diag}}(\frac{1}{8},1,-\frac{1}{4} ,-\frac{1}{9} ,\frac{3}{8},-\frac{1}{2}),\\\nonumber
&&
\hat a^{(1;14)}={\mbox{diag}}(\frac{1}{8} ,-\frac{1}{6} ,\frac{1}{12},-\frac{1}{18},\frac{1}{24},-\frac{1}{12} ).
\end{eqnarray}
In eq.(\ref{ggcc}), we take
\begin{eqnarray}
w_1=1,\;\; w_2=2 ,\;\; c_i =\frac{1}{\pi} I_n  \;\;\Rightarrow \;\;\varkappa^i_\alpha=\frac{3}{2^5},\;\;i=1,2,
\end{eqnarray}
so that it reads
\begin{eqnarray}\label{ggcc_ex}
&&
 g^{(i)}_\alpha(\lambda) =\frac{ \lambda_i}{\pi}
\exp\frac{1}{2}\Big( -\sum_{j=1}^2 \Big((Re\;\lambda_j)^2 + 4 (Im\;\lambda_j)^2 \Big)\Big),\;\;i=1,2.
\end{eqnarray}
Let
\begin{eqnarray}\label{pq}
p_{11}=-i,\;\;p_{12}=- i \,h,\;\;q_{11}=i,\;\;q_{12}= i \,h,
\end{eqnarray}
where $h$ is a positive constant. Then eqs.(\ref{htg3red}) reduce to
\begin{eqnarray}
\tilde g^{(1)}(p_1)= \frac{-2 i}{1+h},\;\;\tilde g^{(2)}(p_1)= \frac{-2 i h}{1+h}.
\end{eqnarray}
Represent $\varepsilon(p_1;t)$ in the form
\begin{eqnarray}
&&
\varepsilon(t)\equiv \varepsilon(p_1;t)={\mbox{diag}}(e^{i X_1},\dots,e^{i X_6}),
\end{eqnarray}
where we introduce variables $X_i$, $i=1,\dots,6$, as the following linear combinations of the variables 
$t_{m_1m_2}$, $m_1,m_2=1,\dots,4$:
\begin{eqnarray}\label{XXX}
&&
X_\alpha = \frac{-2}{1+h}\sum_{m_1,m_2=1}^{4}( \hat a^{(1;1m_1)}_{\alpha}\hat a^{(1;1m_2)}_{\alpha} + 
h \hat a^{(1;1m_1)}_{\alpha+1}\hat a^{(1;1m_2)}_{\alpha+1}) t_{m_1m_2}.
\end{eqnarray}
Since  $\tilde C_{11}= C_{11}^+$ (that follows from comparison of eqs.(\ref{tC1_onesol})),  
we write  eq.(\ref{tVv}) as
\begin{eqnarray}\label{tVv_ex}
&&
\tilde V(t)=
C_0 +
2C_{11}
\hat\theta_0(t)
 C_{11}^+,
 \end{eqnarray}
 where
 \begin{eqnarray}\label{thetav_ex}
&&
\hat\theta_0(t) = \Big(\varepsilon(t) v_0^{-1} \varepsilon^{-1}(t)-
\frac{(1+h)^2 e^{-4(1+h^2)}}{\pi^2} E\Big)^{-1}
 \end{eqnarray}
and $E$ is the $6\times 6$ matrix of units (don't mix it  with the unit matrix!).
Substituting 
values (\ref{pq})  for $p_{1i}$ and $q_{1i}$ into eqs.(\ref{C0}) and (\ref{tC1_onesol}),
we obtain
\begin{eqnarray}\label{C0_H}
C_0&=&-2\sum_{k=1}^2 (\Pi^T)^{k-1}\varkappa^{(k)} \Pi^{k-1}+\\\nonumber
&&
2 \sum_{j=1}^2 \sum_{k=1}^2(\Pi^T)^{k-1}   \varkappa^{(k)}\Pi^{k-1}   p\, p^T (\Pi^T)^{j-1}   \varkappa^{(j)} \Pi^{j-1} = -\frac{3}{8} I_N + \frac{9}{128} E \\\nonumber
C_{11}&=&
\frac{1}{\pi}\sum_{j=1}^2 \sum_{k=1}^2 (\Pi^T)^{k-1} 
\Big(p_{1k}-\varkappa^{(k)}\Pi^{k-1}  p\, p^T 
\sum_{l=1}^K (\Pi^T)^{l-1} p_{1l}\Big) \exp
\Big( -2 \sum_{j=1}^2 p_{1j}^2\Big)\Big)=\\\nonumber
&&
\frac{i e^{-2(1+h^2)}}{\pi}
\Big(- I_N-h \Pi^T + \frac{3 (1+h) }{16 } E\Big).
\end{eqnarray}
Now, let us turn to the constant matrix  $v_0$ appearing in eq.(\ref{thetav_ex}). 
We  consider two particular examples of this matrix and discuss
the particular solutions $U$ of nonlinear PDE associated with them. 

\paragraph{ Example 1.}
As a simple case, let
\begin{eqnarray}\label{v0_ex1}
v_0=\left(\begin{array}{cccccc}
v_1&0&0&0&0&0\cr
0&0&1&0&0&0\cr
0&1&0&0&0&0\cr
0&0&0&1&0&0\cr
0&0&0&0&1&0\cr
0&0&0&0&0&1
\end{array}\right).
\end{eqnarray}
Non-singularity condition (\ref{one22}) requires the following expression for $v_1$:
\begin{eqnarray}
v_1=-5+\frac{(1-d)\pi^2}{(1+h)^2}\exp\Big(4(1+h^2)\Big),\;\;d>0,
\end{eqnarray}
where $d$ is an arbitrary positive parameter.
In this case, solution $U$ depends on the single variable $Z_1=X_3-X_2$. 

Now we fix $h=1/4$.
The absolute values of all elements have oscillating behavior. They may be characterized by the double amplitude  $U^{ampl}_{ij}= |U_{ij}^{max}|-|U_{ij}^{min}|$ and by the average value 
$U^{avr}_{ij}=(|U_{ij}^{max}|+|U_{ij}^{min}|)/2$, which are collected in  Table 1 for three values of $d$:
$d=0.1$, $0.01$ and $0.001$.
Different  shapes of absolute values $|U_{ij}|$ 
 are shown in Fig.\ref{Fig:sol1}a-c.

 \begin{table}[!htb]
\begin{tabular}{|c|c|c||c|c||c|c|}
\hline
      & \multicolumn{2}{c||}{$d=0.1$}&  \multicolumn{2}{c||}{$d=0.01$}  &
      \multicolumn{2}{c|}{$d=0.001$}   \cr
\hline
$|U_{ij}|$ & $U^{ampl}_{ij}$&$U^{avr}_{ij}$ & $U^{ampl}_{ij}$&$U^{avr}_{ij}$ & $U^{ampl}_{ij}$&$U^{avr}_{ij}$\cr
\hline
$|U_{11}|$ & 3.638  &48.201 &269.660  &440.384  &5239.94  &3207.18\cr
$|U_{12}|$ & 0.223  &1.497  & 6.357 & 8.908     & 106.415 &63.6566\cr
$|U_{13}|$ & 0.939  &15.059 & 81.562 &135.842   & 1605.05 &985.039   \cr
$|U_{14}|$ & 1.217  &14.920 & 83.372 & 134.937  & 1608.53 &983.301\cr
$|U_{15}|$ & 1.286  &14.885 & 83.824 &134.710   & 1609.39 &982.867\cr
$|U_{16}|$ & 1.286  &14.879 & 83.824 &134.705   & 1609.39 &982.861\cr
$|U_{22}|$ & 0.010  & 2.416 & 0.127 &2.275      & 2.13798 &1.19072\cr
$|U_{23}|$ & 0.040  & 0.279 &1.905  &2.002      & 32.5786 &18.8045\cr
$|U_{24}|$ & 0.069  & 0.276 &1.960  & 1.993     & 32.6607 &18.7808\cr
$|U_{25}|$ & 0.079  & 0.275 &1.976  &1.990      & 32.6842 &18.7749\cr
$|U_{26}|$ & 0.079  & 0.275 & 1.976  &1.990     & 32.6842 &18.7749\cr
$|U_{33}|$ & 0.229  & 2.387 &24.656  &39.587    & 491.631 &300.228\cr
$|U_{34}|$ & 0.310  & 5.337 & 25.212 &42.299    &492.705  &302.681\cr
$|U_{35}|$ & 0.332  & 5.324 & 25.354 &42.227    &492.975  &302.545\cr
$|U_{36}|$ & 0.332  & 5.324 &25.354  &42.227    &492.975  &302.545\cr
$|U_{44}|$ & 0.406  & 2.311 &25.775  &39.039    &493.774  &299.168\cr
$|U_{45}|$ & 0.430  & 5.281 &25.916  & 41.951   & 494.042 &302.017\cr
$|U_{46}|$ & 0.430  & 5.275 &25.916  &41.945    &494.042  &302.011\cr
$|U_{55}|$ & 0.455  & 2.287 & 26.057 &38.899    &494.309  &298.902\cr
$|U_{56}|$ & 0.455  & 5.269 &26.057  &41.881    &494.309  &301.883\cr
$|U_{66}|$ & 0.455  & 2.287 &26.057  &38.899    &494.309  &298.902\cr
\hline
\end{tabular}
\caption{The amplitudes $U^{ampl}_{ij}$ and the average values $U^{avr}_{ij}$  for  three values of $d$: $d=0.1$, $0.01$ and $0.001$; $h=1/4$ and $v_0$ is given in eq.(\ref{v0_ex1}).}
\label{Table:HPST3}
\end{table}

In Fig.\ref{Fig:sol1}a,
we represent the absolute values $|U_{12}|$ (the upper curve) and $|U_{26}|$ (the lower curve) 
as functions of $Z_1$
for $d=0.1$. Functions $|U_{23}|$, $|U_{24}|$ and $|U_{25}|$  have the  shape of  
$|U_{26}|$ as well, while the absolute values of all other elements have the shape of $|U_{12}|$.

In Fig.\ref{Fig:sol1}b,
we represent the absolute values  $|U_{23}|$ (the big amplitude  curve) and $|U_{22}|$ 
(the small amplitude  curve) as functions of $Z_1$
for $d=0.01$. The absolute values of all other elements have the shape of $|U_{23}|$.

Finally, in Fig.\ref{Fig:sol1}c,
we represent the absolute values $|U_{23}|$ (the upper  curve) and $|U_{22}|$ (the lower  curve)
for $d=0.001$. The absolute values of all other elements have the shape of $|U_{23}|$.

We see from  Table 1, that $|U_{11}|$ has the maximal amplitude   for all three cases.

\begin{figure}
   \epsfig{file=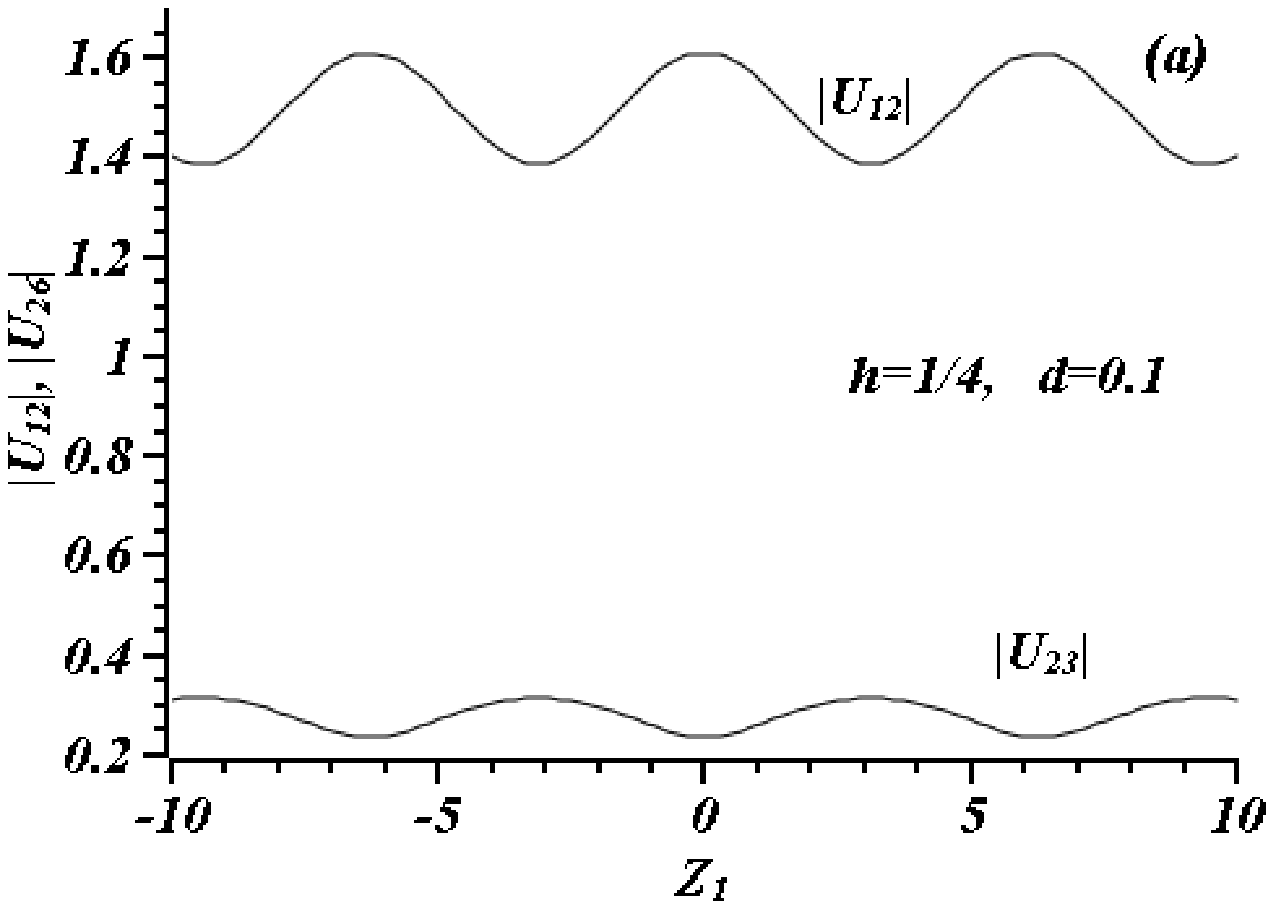, scale=0.7,angle=0}
   \epsfig{file=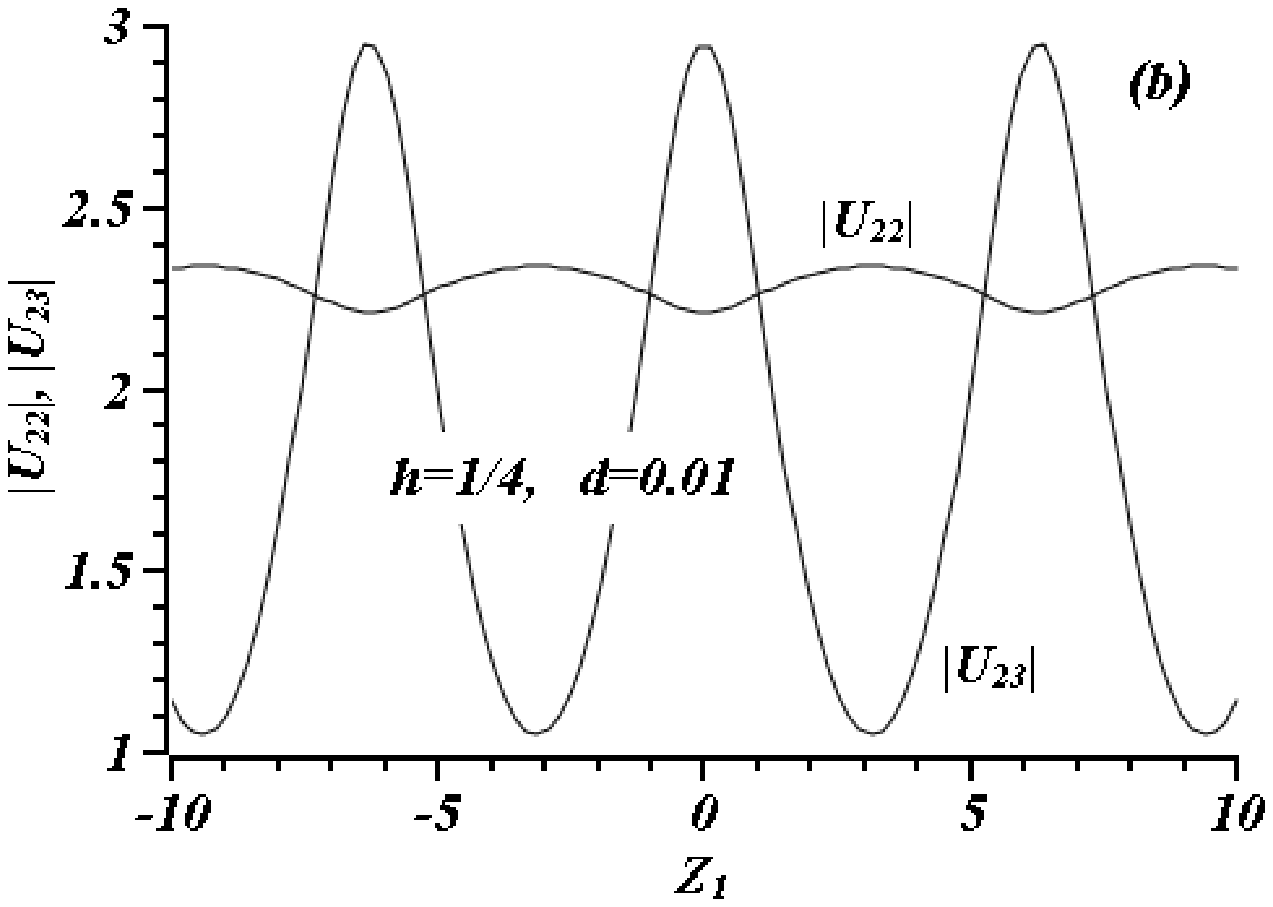, scale=0.7,angle=0}\newline
   \epsfig{file=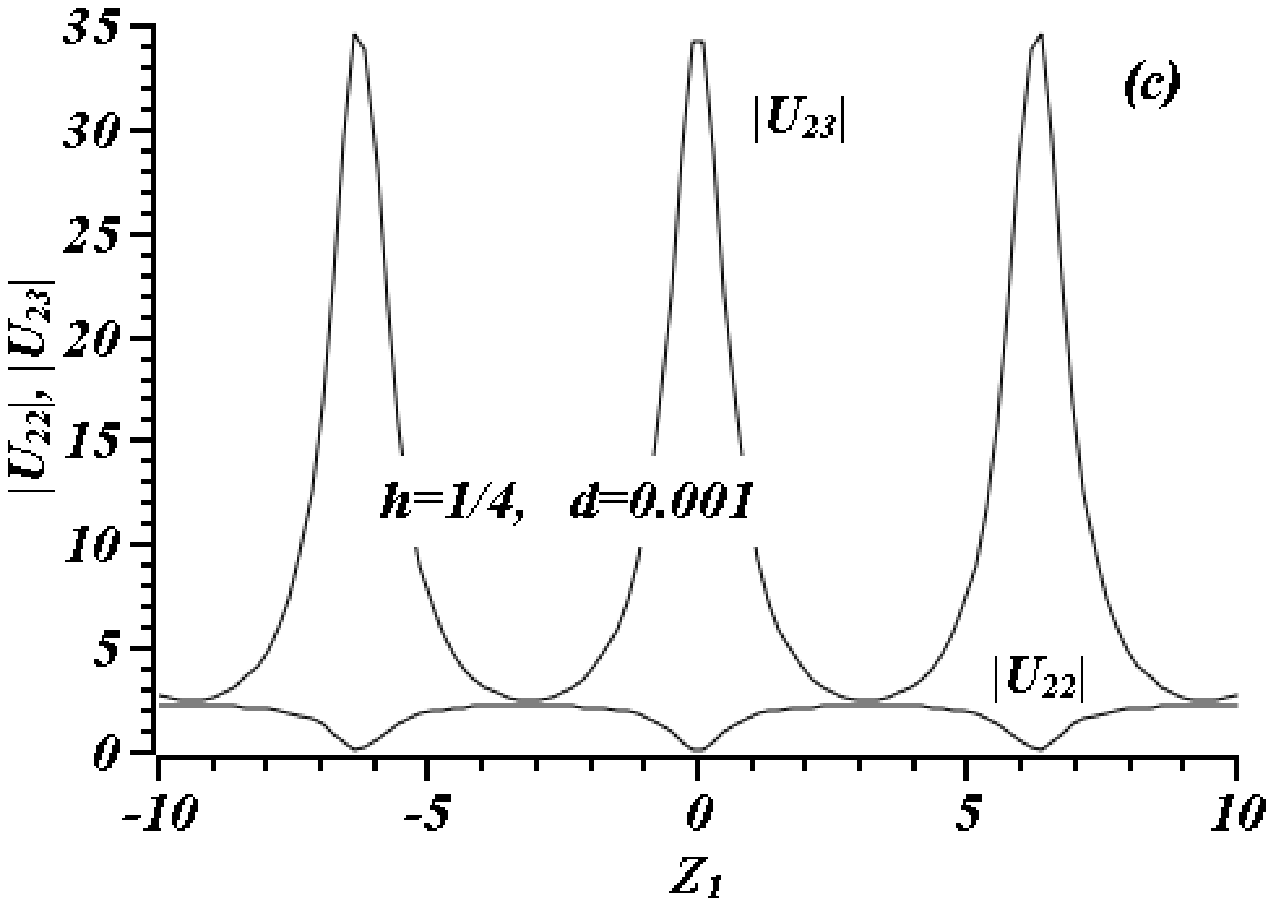, scale=0.7,angle=0}
   \caption{Absolute values of some elements $U_{ij}$ for  $h=1/4$ and $d=0.1,0.01,0.001$ 
   (from the top to the bottom); $v_0$ is given in eq.(\ref{v0_ex1}).}
   \label{Fig:sol1} 
 \end{figure}

 \paragraph{ Example 2.}
 As a more complicated example, we take
\begin{eqnarray}\label{v0_ex2}
v_0=\left(\begin{array}{cccccc}
v_1&1&1&1&1&1\cr
1&0&1&1&1&1\cr
1&1&0&1&1&1\cr
1&1&1&0&1&1\cr
1&1&1&1&0&1\cr
1&1&1&1&1&0
\end{array}\right).
\end{eqnarray}
Condition (\ref{one22}) requires 
\begin{eqnarray}
v_1=-30+\frac{(1-d)\pi^2}{(1+h)^2}\exp\Big(4(1+h^2)\Big),\;\;d>0.
\end{eqnarray}
In this case, solution depends on  five  variables $Z_i=X_{i+1}-X_1$, $i=1,\dots,5$. 
The absolute values of elements  $U_{11}$ and $U_{22}$ as functions of $Z_4$ and $Z_5$ with $h=1/4$, $d=0.001$
(and fixed $Z_1 =Z_2=Z_3=0$ ) are depicted in  Fig.\ref{Fig:sol2}. They are  lattices of lumps. The 
absolute values of all other elements have the shape of $|U_{11}|$.

\begin{figure}
   \epsfig{file=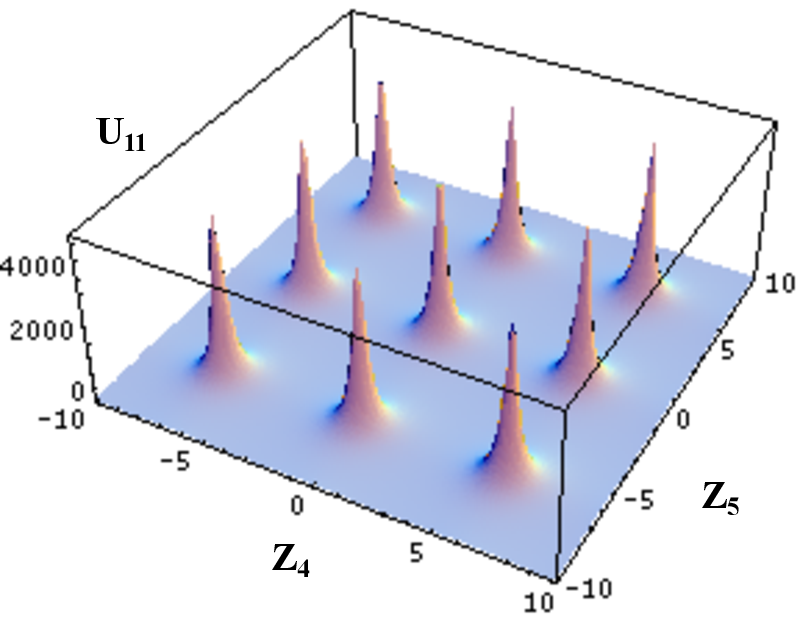, scale=0.8}
   \epsfig{file=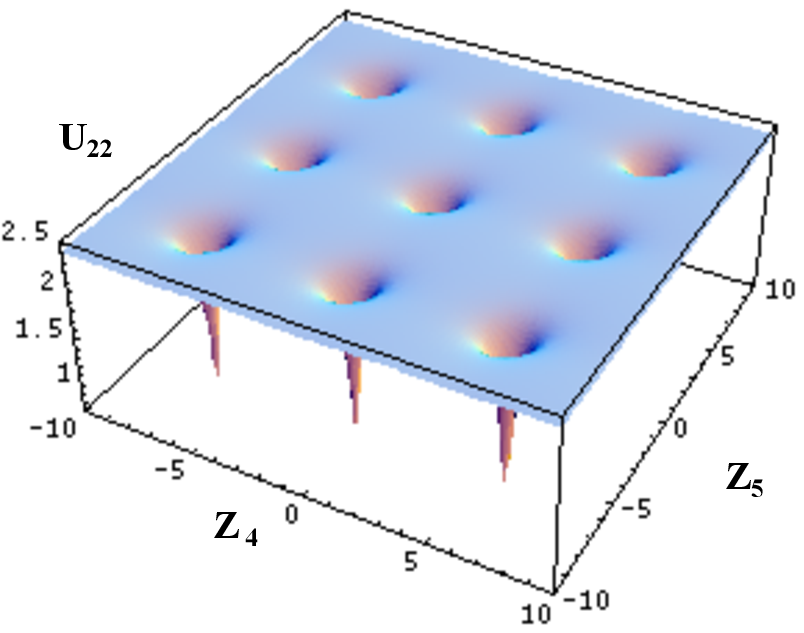, scale=0.8}
   \caption{Elements $U_{11}$ and $U_{22}$ for  $h=1/4$, $d=0.001$; $v_0$ is  given in 
   eq.(\ref{v0_ex2}) and $Z_1 =Z_2=Z_3=0$.}
   \label{Fig:sol2} 
 \end{figure}
Similar to the previous example, the absolute values  $|U_{ij}|$ as functions of $Z_4$ and $Z_5$ 
with fixed $Z_1 =Z_2=Z_3=0$
may be characterized by the double amplitudes $U^{ampl}_{ij}= |U_{ij}^{max}|-|U_{ij}^{min}|$ 
and by the average values $U^{avr}_{ij}=(|U_{ij}^{max}|+|U_{ij}^{min}|)/2$, which are collected in  Table 2.
 \begin{table}[!htb]
\begin{tabular}{|c|c|c|}
\hline
$|U_{ij}|$ & $U^{ampl}_{ij}$&$U^{avr}_{ij}$ \cr
\hline
$|U_{11}|$ &5108.45 &2628.74 \cr
$|U_{12}|$ &91.2786 &48.1621 \cr
$|U_{13}|$ & 1581.04&813.069    \cr
$|U_{14}|$ &1581.04 & 813.069 \cr
$|U_{15}|$ &1580.12 &813.526 \cr
$|U_{16}|$ &1579.89 &813.647 \cr
$|U_{22}|$ &1.61317 &1.58563 \cr
$|U_{23}|$ &28.2704 &14.1456  \cr
$|U_{24}|$ &28.2704 & 14.1399\cr
$|U_{25}|$ &28.211 &14.1696 \cr
$|U_{26}|$ &28.1961 & 14.177\cr
$|U_{33}|$ &489.299 & 249.155\cr
$|U_{34}|$ &489.299  &252.174\cr
$|U_{35}|$ &489.064  &252.297\cr
$|U_{36}|$ &489.006  &252.326\cr
$|U_{44}|$ &489.299 &249.155 \cr
$|U_{45}|$ &489.064 & 252.291\cr
$|U_{46}|$ &489.006 & 252.326\cr
$|U_{55}|$ & 488.726 &249.441\cr
$|U_{56}|$ &488.641 &252.503\cr
$|U_{66}|$ &488.55 & 249.529\cr
\hline
\end{tabular}
\caption{The amplitudes $U^{ampl}_{ij}$ and the average values $U^{avr}_{ij}$  for   $d=0.001$, $h=1/4$;
 $v_0$ is given in eq.(\ref{v0_ex2}). }
\end{table}
 It is remarkable, that if we put to zero any other triad of the
parameters $Z_i$, then we obtain the same shapes for the absolute values $|U_{ij}|$
as functions of two remaining parameters $Z_i$.

\section{Conclusions}
\label{Section:conclusion}

In this paper, we represent a new partially integrable multidimensional 
first-order quasilinear PDE together  with the integral representation of its solution manifold. 
This PDE may be called $n$-wave type equation, although the nonlinear term has 
different structure in comparison with 
the nonlinear term of the well known (2+1)-dimensional completely integrable $n$-wave equation.
The freedom of the solution space is characterized by the arbitrary functions of 
$2(K-1)$   independent variables ($s$-dim.$=2(K-1)$). However the nonlinear PDEs 
are not completely integrable because the increase in the $s$-dimensionality  
causes an increase in the $t$-dimensionality 
of the nonlinear PDE (\ref{S_Q_simple_d_nl_U_h_matr}), 
therewith  $t$-dim. $\sim$ ($s$-dim.$)^2$. 
The $N$-dimensionality  also increases: $N$-dim. $\sim$ $s$-dim.
Thus, the problem of compatible reductions  suppressing  the 
$t$- and/or $N$-dimensionalities of nonlinear PDE is very important. 

We describe the manifold of explicitly representable solutions, 
associated with the case of degenerated kernel of the 
integral operator in (\ref{tV}). The multi-solitary wave solutions are embedded in this manifold.
We derive the condition when such solutions do mot have singularities in the space of independent variables. 
As particular examples, we represent (i) an oscillating  solution depending on a single variable  and (ii) a lamp-lattice 
solution depending on 5 variables.  

We expect that nonlinear equation (\ref{S_Q_simple_d_nl_U_h_matr}) and its 
Hermision reduction (\ref{S_Q_simple_d_nl_U_h}) will be usefull in
study of the multiple-scale expansions of known physical 
systems in hydrodynamics, optics and plasma physics.

Author thanks Prof. P.M.Santini for useful discussions. A part of this work was done during the visit of the 
Institute "La Sapienza" (Roma, Italy) in 2012. This work is partially supported by the Program 
for Support of Leading Scientific Schools (grant No. 3753.2014.2),
and by the RFBR (grant No. 14-01-00389).

\section{Appendix A. Derivation of  nonlinear PDEs
}
\label{Section:appendix}

In this section, we show that  eq.(\ref{S_Q_simple_d_nl_U_h_matr}) can be considered as  
a reduction of a more general equation derivable  via  the dressing method based on the 
following linear  integral equation for the matrix function $W(\lambda;t)$:
\begin{eqnarray}\label{Psi}
&&
 P(\lambda)=W(\mu;t)*\Psi(\mu,\lambda;t)+W(\lambda;t)
 \equiv W(\mu;t)*\Big(\Psi(\mu,\lambda;t)+{\cal{I}}_1(\mu,\lambda)\Big).
\end{eqnarray}
Here
$P(\mu)$, $\Psi(\mu,\lambda;t)$, $W(\lambda;t)$ are the $N\times N$ matrix 
functions of arguments,  $\lambda$, $\mu$, $\nu$ are the complex vector  
parameters (\ref{spectral_par})
of length $K$. 
Function $\Psi$ is the kernel of the integral operator, the star ''$*$''  
means the integration over the space of vector spectral parameter  
defined in eqs.(\ref{def_ast}) and (\ref{def_unit}).
We require that eq.(\ref{Psi}) is uniquely solvable for $W$, i.e., 
the operator $*(\Psi(\mu,\lambda;t)+{\cal{I}}_1(\mu,\lambda))$ is invertible:
\begin{eqnarray}\label{Psi_sol}
W(\lambda;t)= P(\mu)*(\Psi(\mu,\lambda;t)+{\cal{I}}(\mu,\lambda))^{-1}.
\end{eqnarray}
Remember, that we  use independent variables with double indices
and denote 
the whole set of them by $t$: $t=(t_{m_1m_2}: m_1,m_2=1,\dots,D)$.
Let us introduce the dependence on these 
 parameters  through the function $\Psi$, which satisfies  the 
 following system of linear   PDEs with the 
 coefficients independent on $t$:
\begin{eqnarray}\label{S_t}
\Psi_{t_{m_1m_2}}(\lambda,\mu;t)&=&\Big(
B^{(m_1m_2)}(\lambda,\nu)+A(\lambda) C^{(m_1m_2)} P(\nu)\Big)* \Psi(\nu,\mu;t) -\\\nonumber
&&
\Psi(\lambda,\nu;t) *(B^{(m_1m_2)}(\nu,\mu)-A(\nu) C^{(m_1m_2)} P(\mu)),\;\; m_1,m_2=1,\dots,D,
\end{eqnarray} 
where  $B^{(m_1m_2)}(\lambda,\nu)$ and $ A(\lambda)$ are the $N\times N$   
matrix functions of spectral parameters.

\subsection{Derivation of system of compatible linear equations for $W(\lambda;t)$.}
The basic result of this subsection is represented in the following theorem.

{\bf Theorem 1.}
Let matrices $B^{(m_1m_2)}(\lambda,\mu)$ satisfy  the following set of external constraints:
\begin{eqnarray}\label{S_constrain0}
\sum_{m_1=1}^{D} L^{(m_1)} P(\lambda)*(B^{(m_1m_2)}(\lambda,\mu) -A(\lambda) C^{(m_1m_2)} P(\mu))= 0 
,\;\;m_2=1,\dots,D,
\end{eqnarray}
where $L^{(m_1)}$ are some $N\times N$  constant matrices.
Then the
matrix function
$W(\lambda;t)$ obtained as a solution of  integral equation (\ref{Psi}) 
with the kernel $\Psi$ defined by 
eq.(\ref{S_t}) satisfies  the following system of compatible linear equations
\begin{eqnarray}\label{S_lin}
&&
E^{(m_2)}(\lambda;t):=\\\nonumber
&&
\sum_{m_1=1}^{D}L^{(m_1)}\left( W_{t_{m_1m_2}}(\lambda;t) +
 V(t) C^{(m)} W (\lambda;t)
+W(\mu;t) *(B^{(m_1m_2)}(\mu,\lambda)+\right.\\\nonumber
&&
\left.A(\mu)C^{(m_1m_2)} P(\lambda)) \right) =0  ,\;\;
m_2=1,\dots,D,
\end{eqnarray}
where the field $V$ (independent on spectral parameters) is defined as 
\begin{eqnarray}
\label{V}
V(t)=-2 W(\mu;t)*A(\mu) \stackrel{{\mbox{Eq.}}(\ref{Psi_sol})}{=} -2 P*(\Psi+{\cal{I}})^{-1}*A.
\end{eqnarray}

{\bf Proof:} 
To derive eq.(\ref{S_lin}), we 
differentiate eq.(\ref{Psi}) with respect to $t_{m_1m_2}$.  Then, in virtue of eq.(\ref{S_t}), 
one gets the following integral  equation:
\begin{eqnarray}\label{S_lin_1_proof}
&& 
{\cal{E}}^{(m_1m_2)}(\mu;t):=\\\nonumber
&&
P(\nu)*(B^{(m_1m_2)}(\nu,\mu)- P(\nu) C^{(m_1m_2)} A(\mu))=
\tilde E^{(m_1m_2)}(\nu;t)*(\Psi(\nu,\mu;t)+{\cal{I}}_1(\nu,\mu)) ,\\\nonumber
&&
\tilde E^{(m_1m_2)}(\lambda;t)=
W_{t_{m_1m_2}}(\lambda;t) + V(t) C^{(m_1m_2)} W (\lambda;t)
+\\\nonumber
&&
W(\mu;t) *(B^{(m_1m_2)}(\mu,\lambda)+A(\mu)C^{(m_1m_2)} P(\lambda)) 
.
\end{eqnarray}
We consider the following combination of eqs.(\ref{S_lin_1_proof}):
$\sum_{m_1=1}^{D}  L^{(m_1)} {\cal{E}}^{m_1m_2}$. Then, using  external constraints (\ref{S_constrain0}), we  result in:
\begin{eqnarray}\label{S_lin_2}
&& \sum_{m_1=1}^{D} L^{(m_1)} {\cal{E}}^{m_1m_2}:=
\sum_{m_1=1}^{D} L^{(m_1)} \tilde E^{(m_1m_2)}(\nu;t)*(\Psi(\nu,\mu;t)+{\cal{I}}_1(\nu,\mu)) =0
 .
\end{eqnarray}
Since  operator $*(\Psi(\nu,\mu;t)+{\cal{I}}_1(\nu,\mu))$ is invertible, eq.(\ref{S_lin_2}) 
is equivalent to  eq.(\ref{S_lin}).
$\blacksquare$

We refer to constraints (\ref{S_constrain0}) as the external constraints since 
they involve matrices $L^{(m_1)}$ which do not appear in the integral equation 
(\ref{Psi}) as well as in the system of  linear PDEs (\ref{S_t}) defining the function $\Psi$.

System (\ref{S_lin}) is an analogy of the overdetermined system of linear equations in the classical 
inverse spectral transform method. According to that method, the system of 
nonlinear PDEs for the potentials of the overdetermined linear system appears as 
the compatibility condition of this linear system. However, the  nonlinear PDE for the matrix field $V$
may not be obtained as the compatibility condition  in our case because of the  term $ W(\mu;t) *(B^{(m_1m_2)}(\mu,\lambda)+A(\mu)C^{(m_1m_2)} P(\lambda))$   
in  eq.(\ref{S_lin}). 
Therefore, we represent another algorithm of derivation of nonlinear PDE in the next subsection.

\subsection{Derivation of the first order nonlinear PDE for the field $V(t)$}
\label{Section:S_first_order}
{\bf Theorem 2.} In addition to  eqs.(\ref{Psi},\ref{S_t}) and  external constraints (\ref{S_constrain0}), we impose another set of   external constraints:
\begin{eqnarray}\label{S_constrain1}
&&
\sum_{m_2=1}^{D}( B^{(m_1m_2)}(\lambda,\nu)+ A(\lambda) C^{(m_1m_2)} P(\nu) )* A(\nu) R^{(m_2)} = 0,\\\nonumber
&&
m_1=1,\dots,D,
\end{eqnarray}
where $R^{(m_2)}$ are  some $N\times N$ constant matrices.
Then the $N\times N$ matrix function $V(t)$ is a  solution to the  following nonlinear PDE:
\begin{eqnarray}\label{S_Q}
&&
\sum_{m_1,m_2=1}^{D} L^{(m_1)}\left( V_{t_{m_1m_2}} + V C^{(m_1m_2)} V
\right)R^{(m_2)} =0.
\end{eqnarray}

{\bf Proof:} 
Applying the operator  $*(-2  A)$ to eq.(\ref{S_lin}) from the right, one gets the following 
equation
\begin{eqnarray}\label{S_2U}
&&
E^{(m_2)}(t)=E^{(m_2)}(\lambda;t)*  A(\lambda):=\\\nonumber
&&
\sum_{m_1=1}^{D} L^{(m_1)}\left( V_{t_{m_1m_2}} + VC^{(m_1m_2)} V
+U^{(m_1m_2)}  \right) =0 ,
\end{eqnarray}
which introduces a new set of fields $U^{(m_1m_2)}$, $m_1,m_2=1,\dots,D$,
\begin{eqnarray}
\label{UW}
U^{(m_1m_2)}(t)=-2 W(\mu;t) *(B^{(m_1m_2)}(\mu,\nu)+A(\mu)C^{(m_1m_2)} P(\nu))*A(\nu).
\end{eqnarray}
Due to the external constraints (\ref{S_constrain1}), we may eliminate these fields using the
appropriate combinations of eqs.(\ref{S_2U}). Namely, the combination 
$\sum_{m_2=1}^{D_2} E^{(m_2)} R^{(m_2)}$ results in  system 
(\ref{S_Q}). $\blacksquare$

Nonlinear PDE (\ref{S_Q}) is the general form of  PDEs considered below. 
First it was derived in 
\cite{Z2010}. However, the acceptable structure of
the constant matrix  coefficients $L^{(m_1)}$, $R^{(m_2)}$
and $C^{(m_1m_2)}$  as well as the richness 
of the solution space have not been investigated to the full extent in that reference.

Below we show that the structure of the coefficients $L^{(m_1)}$, $R^{(m_2)}$ and $C^{(m_1m_2)}$ is 
defined by  the solution $\Psi$  of  system of linear PDEs (\ref{S_t}). In Sec. \ref{Section:Psi}, 
choosing the special form for the constant matrices $C^{(m_1m_2)}$, we represent a family of 
solutions to the system of linear PDEs (\ref{S_t}) leading to the multidimensional 
PDE of non-classical type. Another choice of the constant matrices $C^{(m_1m_2)}$, leading to the classical
(2+1)-dimensional $n$-wave equation, is considered in Sec.\ref{Section:appendixD}.

\subsection{Construction of  kernel $\Psi$ as a solution to   system of  linear PDEs
(\ref{S_t}). Special form of  matrices $C^{(m_1m_2)}$}
\label{Section:Psi}
In this section, we denote $m=(m_1m_2)$ for the sake of brevity. Accordingly, 
$\sum_m \equiv \sum_{m_1,m_2=1}^D$. 
In order to construct  solutions to  nonlinear PDE (\ref{S_Q}), 
we have to find the explicit form of the function $\Psi$ solving  
 system of linear PDEs (\ref{S_t}). Let us look for the  solution  $\Psi$ in the following form:
\begin{eqnarray}\label{Chi_sol}
\Psi(\lambda,\mu) =  \chi(\lambda,\nu) *\Big(\varepsilon(\nu;t) {\cal{C}}(\nu,\tilde\nu) 
\tilde \varepsilon(\tilde\nu,t)\Big)*
\tilde \chi(\tilde\nu,\mu) ,
\end{eqnarray}
where 
\begin{eqnarray}\label{I_structure}\label{I_chi1}
&&\varepsilon(\nu;t)=e^{\sum_m T^{(m)}(\nu) t_m} ,\;\;\;\tilde \varepsilon(\nu;t)=e^{-\sum_m \tilde T^{(m)}(\nu) t_m} .
\end{eqnarray}
Here $\chi$ and $\tilde \chi$ are the  $N\times N$ invertible  matrix operators, $T^{(m)}$ and $\tilde T^{(m)}$ are the diagonal $N\times N$ matrix functions of the spectral parameter.  
Substituting function $\Psi$ given in eq.(\ref{Chi_sol}) into eq.(\ref{S_t})
we obtain
\begin{eqnarray}
&&
 \Big(\chi T^{(m)} - (B^{(m)} +  A C^{(m)} P)*\chi\Big)*\varepsilon*C*\tilde \varepsilon*
\tilde \chi -\\\nonumber
&&
\chi \varepsilon*C*\tilde \varepsilon*\Big(\tilde T^{(m)} \tilde \chi - \tilde \chi * 
(B^{(m)}-  A C^{(m)} P) \Big)=0.
\end{eqnarray}
Each of two terms in this equation must be identical to zero, 
which suggests us 
 the following two equations relating  $C^{(m)}$, 
$B^{(m)}$, $T^{(m)}$ and $\tilde T^{(m)}$:
\begin{eqnarray}\label{eqA}
&&\chi T^{(m)} - (B^{(m)} +  A C^{(m)} P)*\chi=0,\\\label{eqB}\label{eqB2}
&&\tilde T^{(m)} \tilde \chi - \tilde \chi *( B^{(m)} -A C^{(m)} P)=0.
\end{eqnarray}
Solving  eq.(\ref{eqA})  for   $B^{(m)}$, we obtain:
\begin{eqnarray}\label{Br0}\label{Br}
B^{(m)} = (\chi T^{(m)})*\chi^{-1} -
 A C^{(m)} P,
\end{eqnarray}
which   defines the operator $B^{(m)}$.
Substituting this expression into  eq.(\ref{eqB2}) and 
applying  the operator $* \chi$ from the right side, we result in the following relation:
\begin{eqnarray}\label{rBr}
 R(\lambda,\mu) T^{(m)}(\mu)-\tilde T^{(m)}(\lambda) R(\lambda,\mu)  =2 r(\lambda) C^{(m)}  \tilde r(\mu),\;\;\forall m,
\end{eqnarray} 
where we introduce notations
\begin{eqnarray}\label{R_r_r}
R(\lambda,\mu)=\tilde \chi*\chi,\;\;\;
r(\lambda)=\tilde\chi(\lambda,\nu)*A(\nu)
,\;\;\;
\tilde r(\mu)= P(\nu)*\chi(\nu,\mu).
\end{eqnarray}
Thus, we have to find  functions  $R$, $r$, $\tilde r $, $T^{(m)}$, $\tilde T^{(m)}$ 
and constant matrices $C^{(m)}$  satisfying eqs.(\ref{rBr}).

Remark, that 
there is a particular solution to  system (\ref{rBr}) leading to the classical integrable (2+1)-dimensional
$n$-wave equation, which is considered in  Appendix D, Sec.\ref{Section:appendixD}. 
Here we study another solution to  system (\ref{rBr})  resulting in a 
new multidimensional first order quasilinear  equation.

In order to solve eq.(\ref{rBr}), we propose the following form of the constant matrix $C^{(m)}$: 
\begin{eqnarray}\label{Cm}
&&
C^{(m)} = \xi \xi^{(m)} -\eta^{(m)}  \eta 
\end{eqnarray}
and take the  function $R(\lambda,\mu)$  in the form 
\begin{eqnarray}\label{R_alphabeta}
&&
R(\lambda,\mu)= {\cal{I}}(\lambda,\mu) + r(\lambda) \xi \eta \tilde r(\mu).
\end{eqnarray}
Here $\xi$ and $\eta^{(m)}$ are $N\times 1$ constant matrices, while
$\xi^{(m)}$ and $\eta$ are $1\times N$ constant  matrices.
Substituting eqs.(\ref{R_alphabeta}) and (\ref{Cm}) into 
 eq.(\ref{rBr}) we obtain:
\begin{eqnarray}\label{rBr_sep}
&&
{\cal{I}}(\lambda,\mu) ( T^{(m)}(\mu)  - \tilde T^{(m)}(\lambda) )  +
r(\lambda) \xi \eta \tilde r(\mu) T^{(m)}(\mu)  - 
 \tilde T^{(m)}(\lambda)r(\lambda) \xi  \eta \tilde r(\mu)  =\\\nonumber
&&
2 r(\lambda) \xi \xi^{(m)} \tilde r(\mu)-2 r (\lambda) 
 \eta^{(m)}  \eta \tilde r(\mu) .
\end{eqnarray}
Eq.(\ref{rBr_sep}) may be splitted into  three following relations:
\begin{eqnarray}\label{rBr_1}
&&
\tilde T^{(m)}(\lambda)  =  T^{(m)}(\lambda),\\\label{rBr_3}
&&
r(\lambda)\eta^{(m)}=\frac{1}{2}   T^{(m)}(\lambda)r(\lambda) \xi,\\\label{rBr_2}
&&
 \xi^{(m)}\tilde r(\mu)=\frac{1}{2}\; \eta \tilde r(\mu) T^{(m)}(\mu) .
\end{eqnarray}
Owing to eq.(\ref{rBr_1}), we have 
\begin{eqnarray}\label{tvarepsilon}
\tilde\varepsilon(\lambda;t) = \varepsilon^{-1}(\lambda;t).
\end{eqnarray}
Let us analyze eqs.(\ref{rBr_3}) and (\ref{rBr_2}).
Solving eq.(\ref{rBr_3}) for $T^{(m)}$, we obtain:
\begin{eqnarray}\label{TT}
T^{(m)}_{\alpha}(\lambda) = 2 \frac{ \sum_{\gamma=1}^N   r_{\alpha\gamma}(\lambda) \eta^{(m)}_{\gamma 1}}{
\sum_{\gamma=1}^N  r_{\alpha\gamma}(\lambda)  \xi_{\gamma 1} }.
\end{eqnarray}
In other words, we relate the $\lambda$-dependence of the diagonal 
elements of $T^{(m)}(\lambda)$ with the $\lambda$-dependence of the 
elements of the matrix $r(\lambda)$. 
Substituting eq.(\ref{TT}) into eq.(\ref{rBr_2}) we obtain 
\begin{eqnarray}\label{rBr_comp}
  \frac{ \sum_{\gamma=1}^N   r_{\alpha\gamma}(\lambda) \eta^{(m)}_{\gamma 1}}{
\sum_{\gamma=1}^N  r_{\alpha\gamma}(\lambda)  \xi_{\gamma 1} }=  
 \frac{ \sum_{\gamma=1}^N  \tilde  r_{\gamma \alpha}(\lambda) \xi^{(m)}_{1\gamma }}{
\sum_{\gamma=1}^N  \tilde r_{\gamma\alpha}(\lambda)  \eta_{1\gamma } },
\end{eqnarray}
which relates elements of $r$ and $\tilde r$. In particular, relation (\ref{rBr_comp}) becomes an identity if 
equation (\ref{rBr_2}) is the transposition  of  eq. (\ref{rBr_3}), i.e.,
\begin{eqnarray}\label{H1}
\tilde r(\lambda) = r^T(\lambda),\;\; \eta = \xi^T,\;\; \xi^{(m)} = ( \eta^{(m)})^T.
\end{eqnarray}
Thus, to satisfy eq.(\ref{rBr}), we  use the matrices $C^{(m)}$ and $R$
given by, respectively,  eq.(\ref{Cm}) and eq.(\ref{R_alphabeta}), 
identify $\tilde T^{(m)} = T^{(m)}$ (eq.(\ref{rBr_1})), 
relate $T^{(m)}(\lambda)$ with $r(\lambda)$  by eq.(\ref{TT}) and impose  additional relations 
(\ref{H1}). 

Now, substituting eqs.(\ref{Cm}) and (\ref{H1}) in
eq.(\ref{S_Q}), we obtain the following nonlinear PDE
\begin{eqnarray}\label{S_Q_H1}\label{S_Q_simple}
&&
\sum_{m_1,m_2=1}^{D} L^{(m_1)}\left( V_{t_{m_1m_2}} + V \xi (\eta^{(m_1m_2)})^T  V -
V \eta^{(m_1m_2)}  \xi^T  V
\right)R^{(m_2)} =0.
\end{eqnarray}

Finally, using
formulas (\ref{Chi_sol}) and (\ref{R_r_r}), we write expression (\ref{V}) for $V$   as follows:
\begin{eqnarray}\label{V2}
V(t)&=&-2  r^T(\lambda) *(\Psi_0(\lambda,\mu;t)*R(\mu,\nu)+{\cal{I}}(\lambda,\nu))^{-1}*R^{-1}(\nu,\tilde\nu)*r(\tilde\nu),
\\\label{Psi00}
&&
\Psi_0(\lambda,\mu;t) = \varepsilon(\lambda;t) {\cal{C}}(\lambda,\mu)  \varepsilon^{-1}(\mu;t).
\end{eqnarray}

Let us  give another representation for $V$  using
the explicit expression for the  inverse of the operator  $R$,
\begin{eqnarray}
\label{Rinv}
&&
R^{-1}(\lambda,\mu)=  {\cal{I}}(\lambda,\mu) -
 \frac{r(\lambda) \xi \xi^T r^T(\mu) }{1+Q} ,
\\\label{Q}
&&
Q =\xi^T r^T(\lambda)*r(\lambda) \xi.
\end{eqnarray}
 Substituting this expression in combinations 
 $ r^T *R^{-1}$ and $R^{-1}*r$ 
 we obtain
\begin{eqnarray}\label{Gamma}
 r^T(\nu) *R^{-1}(\nu,\lambda) &=&
 \Big(1 - \frac{r^T*r \xi \xi^{T}}{1+Q}\Big)  r^T(\lambda)=
\Gamma r^T(\lambda),\\\label{GammaT}
R^{-1}(\lambda,\nu) * r(\nu)& =& 
r(\lambda)\Big(1- \frac{\xi  \xi^T r^T * r}{1+Q} \Big)=r(\lambda)\Gamma^T,
\end{eqnarray}
where 
\begin{eqnarray}
\label{Gamma2}
&&
\Gamma= 1 - \frac{r^T*r \xi \xi^T}{1+Q}.
\end{eqnarray}
Finally, using eqs.(\ref{Gamma}) and (\ref{GammaT}), we transform eq.(\ref{V2}) into the following one:
\begin{eqnarray}
\label{Vp}
V&=&-2 \Gamma r^T(\lambda)*R(\lambda,\mu)*
(\Psi_0(\mu,\tilde \mu;t)* R(\tilde \mu,\nu) +{\cal{I}} (\mu,\nu))^{-1}*r(\nu) \Gamma^T.
\end{eqnarray}
We emphasize, that $N\times 1$  matrix $r^T*r \xi$  in  definition of $\Gamma$ (\ref{Gamma2}) must be finite, i.e.
\begin{eqnarray}\label{rr_inf}
|(r^T*r \xi)_{\alpha 1}|<\infty,\;\;\alpha=1,\dots,N.
\end{eqnarray}

\subsection{External constraints (\ref{S_constrain0}) and (\ref{S_constrain1})}
\label{Section:constraints}
Now we have to satisfy  external constraints  (\ref{S_constrain0}) and  (\ref{S_constrain1}). 
First,  using 
the substitution $(B^{(m)}-AC^{(m)}P)=\tilde \chi^{-1} \tilde T^{(m)}\tilde \chi$ (which follows from 
eq.(\ref{eqB2})), we transform  constraint  (\ref{S_constrain0}) to
 the following form
\begin{eqnarray}
\label{T1}
\sum_{m_1}L^{(m_1)}
P*(\tilde  \chi^{-1}\tilde T^{(m)})*\tilde  \chi
=0.
\end{eqnarray}
Or, applying $*\tilde \chi^{-1}$ and using notations (\ref{R_r_r}) with relations (\ref{rBr_1},\ref{H1}), we write
eq.(\ref{T1}) as  
\begin{eqnarray}
\label{T1_2}
\sum_{m_1}L^{(m_1)}
 r^T*R^{-1} T^{(m)}
=0.
\end{eqnarray}
In a similar way, we transform   constraint (\ref{S_constrain1}) 
using  substitution  $(B^{(m)} +  A C^{(m)} P )= \chi*(T^{(m)} \chi^{-1})$
 (following from eq.(\ref{eqA})) with notations (\ref{R_r_r}) and applying $\chi^{-1}*$.
As a result we obtain
\begin{eqnarray}
\label{T2_2}
\sum_{m_2}   T^{(m)}  R^{-1} * r R^{(m_2)}=0.
\end{eqnarray}
 
Now, substituting eqs.(\ref{Gamma}) and (\ref{GammaT})  into constraints  (\ref{T1_2}) and (\ref{T2_2}),
we obtain
\begin{eqnarray}\label{C1}
&&\sum_{m_1=1}^{D} L^{(m_1)} \Gamma r^T(\lambda) T^{(m)}(\lambda) =0,\;\;m_2=1,\dots,D,\\\label{C2}
&&
\sum_{m_2=1}^{D}T^{(m)}(\lambda) r(\lambda) \Gamma^T R^{(m_2)} =0, \;\;m_1=1,\dots,D,
\end{eqnarray}
where $T^{(m)}$ is related with $r$ by eq.(\ref{TT}).
For convenience, we  introduce the notation ${\mbox{diag}} \, A$ for the diagonal matrix with the diagonal elements $A_{\alpha 1}$, where $A$ is $N\times 1$ matrix. 
Let us substitute  $T^{(m)}$ from eq.(\ref{TT}) into  eqs.(\ref{C1},\ref{C2}) 
and multiply the result by the non-degenerate diagonal matrix ${\mbox{diag}} (r*\xi)$
from the right and left sides respectively. Finally we obtain the external constraints in the following form:
\begin{eqnarray}\label{C1_M}
&&
\sum_{m_1=1}^{D} L^{(m_1)} \Gamma r^T(\lambda) \;{\mbox{diag}}\Big(
 r(\lambda) \eta^{(m_1m_2)}\Big) =0,\;\;m_2=1,\dots,D,
 \\\label{C2_M}
&&
\sum_{m_2} {\mbox{diag}}\Big(
 r(\lambda) \eta^{(m_1m_2)}\Big) r(\lambda) \Gamma^T R^{(m_2)} =0, \;\;m_2=1,\dots,D,
\end{eqnarray}
Eqs.(\ref{C1_M},\ref{C2_M}) represent the  system  of  nonlinear equations for the elements of $r$
involving constant (non-diagonal in general) matrices $L^{(m_1)}$, $R^{(m_2)}$.

\subsubsection{Resolving  external constraints (\ref{C1_M}) and (\ref{C2_M}) using a special form of 
 function $r(\lambda)$}
Both constraints (\ref{C1_M}) and (\ref{C2_M}) depend on the spectral parameter $\lambda$ through the function 
 $r(\lambda)$. They can be resolved considering 
$r(\lambda)$ as a linear combination of  $K$ arbitrary diagonal  functions  
$ g^{(i)}(\lambda)$ ($i=1,\dots,K$) of the spectral parameter
 $\lambda$, i.e.,
\begin{eqnarray}
\label{g}
&&
r(\lambda)= \sum_{j=1}^K g^{(j)}(\lambda) a^{(j)}.
\end{eqnarray}
Substituting  eq.(\ref{g})  
in eqs.(\ref{C1_M}) and (\ref{C2_M})  we obtain:
\begin{eqnarray}\label{C1_M21}
&&
\sum_{{i,j=1}\atop{j\ge i}}^K  Z^{(m_2; ij)} g^{(i)}(\lambda)  g^{(j)}(\lambda)  =0,
\\\label{C2_M21}
&&
\sum_{{i,j=1}\atop{j\ge i}}^K  g^{(i)}(\lambda)  g^{(j)}(\lambda)  \tilde Z^{(m_1; ij)} =0,
\end{eqnarray}
where 
\begin{eqnarray}\label{Z1gen}
Z^{(m_2;ii)}&=&\sum_{m_1=1}^{D} L^{(m_1)}
 \hat a^{(i)}  \hat a^{(i;m_1m_2)},\\\label{Z2gen}
Z^{(m_2;ij)}&=&\sum_{perm(i,j)}\sum_{m_1=1}^{D}L^{(m_1)}
 \hat a^{(i)} \hat  a^{(j;m_1m_2)},\;\;i\neq j\\
 \label{Z3gen}
\tilde Z^{(m_1;ii)}&=&\sum_{m_1=1}^{D} 
  \hat a^{(i;m_1m_2)} (\hat a^{(i)})^T R^{(m_2)} ,\\\label{Z4gen}
\tilde Z^{(m_1;ij)}&=&\sum_{perm(i,j)}\sum_{m_1=1}^{D}
  \hat  a^{(j;m_1m_2)}(\hat a^{(i)})^TR^{(m_2)},\;\;i\neq j,
 \end{eqnarray}
and
\begin{eqnarray}\label{hatai}
 \hat a^{(i)} =  \Gamma (a^{(i)})^T\;\;\;\Rightarrow \;\;\; a^{(i)} =( \hat a^{(i)})^T (\Gamma^T)^{-1} .
\end{eqnarray}
Assuming the  linear independence of $g^{(i)}(\lambda)$, $i=1,\dots,K$, 
we conclude that eqs.(\ref{C1_M21},\ref{C2_M21}) hold if 
\begin{eqnarray}\label{Z}
&&Z^{(m_2;ij)}=0,\;\;m_2=1,\dots,D,\\\label{tZ}
&&
\tilde Z^{(m_1;ij)}=0,\;\;m_1=1,\dots,D, \;\;i,j=1,\dots,K.
\end{eqnarray}
In other words,  system of equations (\ref{C1_M21},\ref{C2_M21}) depending on the spectral parameter 
is equivalent to  system (\ref{Z},\ref{tZ}) independent on the 
spectral parameter. This system will be solved below.

\subsubsection{Structure of operators $T^{(m_1m_2)}$ in (\ref{TT}) associated with  $r(\lambda)$ given in (\ref{g})}

Let us show that representation (\ref{g}) leads to  ($K-1)$ arbitrary 
functions of spectral parameters in the exponent $\varepsilon(\lambda)$, 
see eqs.(\ref{I_chi1}).   
In fact,
substituting eq.(\ref{g}) into  eq.(\ref{TT}) one gets:
\begin{eqnarray}\label{I_mT}
T^{(m_1m_2)}(\lambda) =\sum_{j=1}^K 
\tilde g^{(j)}(\lambda) \hat a^{(j;m_1m_2)},
\end{eqnarray}
where the elements of the diagonal matrices $\tilde g^{(j)}$ are defined by the formulas
\begin{eqnarray}\label{tg}
\tilde g^{(j)}_\alpha(\lambda) = \frac{2 g^{(j)}_\alpha(\lambda)}{\sum_{i=1}^K \sum_{\gamma=1}^N
 g^{(i)}_\alpha(\lambda) a^{(i)}_{\alpha\gamma}\xi_{\gamma1}},
\end{eqnarray}
and the elements of the diagonal matrices $\hat a^{(j;m_1m_2)}$ are defined as
\begin{eqnarray} \label{I_hata}
 \hat  a^{(j;m_1m_2)}_\alpha =\Big(a^{(j)}\eta^{(m_1m_2)}\Big)_{\alpha1},\;\;
 \alpha=1,\dots,N,\;\;m_1,m_2=1,\dots,D,\;\;i=1,\dots,K
\end{eqnarray}
with 
\begin{eqnarray}\label{norm}\label{I_norm}
\hat a^{(i;1 1)}_\beta =1,\;\;i=1,\dots,K,\;\;\beta=1,\dots,N.
\end{eqnarray}

Eq.(\ref{I_mT}) shows that all $T^{(m_1m_2)}$ are linear combinations of $K$ diagonal 
functions $\tilde g^{(j)}(\lambda)$, $j=1,\dots,K$, and each of these functions 
involves $(K-1)$ arbitrary functions $\hat g^{(j)}(t)$ of spectral parameters, given in (\ref{hatgg}),
as follows:
\begin{eqnarray}\label{htg}
&&
\tilde g^{(1)}_\alpha(\lambda) = 2 \left(  \sum_{\gamma=1}^Na^{(1)}_{\alpha\gamma}\xi_{\gamma1} + 
\sum_{i=2}^K \sum_{\gamma=1}^N
\hat g^{(i)}_\alpha(\lambda) a^{(i)}_{\alpha\gamma}\xi_{\gamma1}\right)^{-1}
,\\\nonumber
&&
\tilde g^{(j)}_\alpha(\lambda) =  \hat g^{(j)}_\alpha(\lambda)\tilde g^{(1)}_\alpha(\lambda).
\end{eqnarray}
 Thus, $\tilde g^{(k)}(\lambda)$ are parametrized by  $(K-1)$ arbitrary functions  
 $\hat g^{(i)}(\lambda)$   ($i=2,\dots, K$) of the vector spectral parameter $\lambda$.
 Consequently, the same holds for the  functions $T^{(m_1m_2)}(\lambda)$.

\subsection{Solution of system (\ref{Z},\ref{tZ})
}
 Hereafter we consider the case of non-degenerate  matrices 
 $\hat a^{(i;m_1m_2)}$. 
First of all, 
we  reduce the number of equations  in  system (\ref{Z},\ref{tZ}) 
decomposing each of  the constant diagonal  matrices  $\hat a^{(i;m_1m_2)}$ 
into the  pair of other  diagonal matrices 
$\hat a^{(i;m_1 1)}$ and $\hat a^{(i;1 m_2)}$ as follows:
\begin{eqnarray}\label{I_aa_gen}
\hat a^{(i;m_1 m_2)} =\hat a^{(i;m_1 1)} \hat a^{(i;1 m_2)},
\;\;\; m_1,m_2=1,\dots,D,\;\; i=1,\dots,K,
\end{eqnarray}
with normalization (\ref{norm}).
Substituting decomposition (\ref{I_aa_gen}) into constraints (\ref{Z}) and (\ref{tZ}) we recognize
that both are satisfied if, respectively, 
\begin{eqnarray}
\label{I_Z2gen21f}
&&
\sum_{m_1=1}^{D}L^{(m_1)}
 \hat a^{(i)} \hat  a^{(j;m_11)}=0,\;\;i,j=1,\dots,K
\end{eqnarray}
and
\begin{eqnarray}
\label{tZ2gen2}\label{I_tZ1gen21}\label{I_tZ2gen22}
 \sum_{m_2=1}^{D} 
   \hat a^{(i;1m_2)}(\hat a^{(j)})^TR^{(m_2)}=0,\;\;i,j=1,\dots,K.
\end{eqnarray}
We obtain the general solution of system (\ref{I_Z2gen21f},\ref{tZ2gen2}) in Sec.\ref{Section:redZ}.
A reduced form of this system is derived in Sec.\ref{Section:DlessN}.

\subsubsection{General solution of external constraints 
(\ref{I_Z2gen21f},\ref{I_tZ2gen22}): relations among constant matrices in  
nonlinear PDE (\ref{S_Q_simple})}
\label{Section:redZ}

System (\ref{I_Z2gen21f}) consists of $K^2$ matrix $N\times N$ equations and can be  
considered  as a system solvable for the constant matrices 
$L^{(m_1)}$, $m_1=1,\dots,D$. Similarly the system of $K^2$ equations (\ref{tZ2gen2}) can  be solved for the  constant matrices $R^{(m_2)}$, $m_2=1,\dots,D$.
 In order to avoid additional constraints on the  matrices $\hat a^{(i;m_11)}$,
 $\hat a^{(i;1m_2)}$ and $\hat a^{(i)}$, we require
\begin{eqnarray}\label{DK2}
 D= K^2+1.
 \end{eqnarray}
{{} Then system (\ref{I_Z2gen21f}) can be considered as a system for $D-1$ matrices 
$L^{(m_1)}$, $m_1=2,\dots, D$, with the arbitrary matrix $L^{(1)}$,  
while   system (\ref{tZ2gen2}) may be considered as a system for the matrices 
$R^{(m_2)}$, $m_2=2,\dots, D$ with the arbitrary matrix $R^{(1)}$. Both $L^{(1)}$ and $R^{(1)}$ can be 
identity matrices without the loss of generality.}

Now we determine the  $N$-dimensionality 
of the nonlinear PDE (\ref{S_Q_simple}) required for resolvability of 
relations ({\ref{I_aa_gen}). 
Remember, that matrices $\hat a^{(i;m_1m_2)}$ 
must satisfy their definitions (\ref{I_hata}), 
which, in view of eq.(\ref{I_aa_gen}), take the form 
\begin{eqnarray} \label{hata}
 &&
 \hat  a^{(j;m_11)}_\alpha \hat  a^{(j;1m_2)}_\alpha =\Big(a^{(j)}\eta^{(m_1m_2)}\Big)_{\alpha1},
 \;\;\hat a^{(j;11)}_\alpha=1,
 \\\nonumber
 &&
 \alpha=1,\dots,N,\;\;m_1,m_2=1,\dots,D,\;\;i=1,\dots,K.
\end{eqnarray}
Let us consider  system (\ref{hata}) as a system for  $a^{(i)}$ and
write it in a  matrix form as follows.
First, we note that  the number of different $\hat a^{(i;m_1m_2)}$ for any $i$ equals $D^2$. 
Next, we introduce the $N\times D^2$  matrices $\hat \xi$ and $\hat A^{(i)}$,
\begin{eqnarray}\label{matr_xi0}
\hat\xi&=&\left(
\begin{array}{cccccccccc}
\eta^{(11)}_{1 1} &\cdots &\eta^{(1D)}_{1 1}&\eta^{(21)}_{1 1}&\cdots& \eta^{(2D)}_{1 1}&\cdots
&\eta^{(D1)}_{1 1}&\cdots& \eta^{(DD)}_{1 1}\cr
\eta^{(11)}_{2 1} &\cdots &\eta^{(1D)}_{2 1}&\eta^{(21)}_{2 1}&\cdots& \eta^{(2D)}_{2 1}&\cdots
&\eta^{(D1)}_{2 1}&\cdots& \eta^{(DD)}_{2 1}\cr
\cdots&\cdots&\cdots&\cdots&\cdots&\cdots&\cdots&\cdots&\cdots&\cdots\cr
\eta^{(11)}_{N 1} &\cdots &\eta^{(1D)}_{N 1}&\eta^{(21)}_{N 1}&\cdots& \eta^{(2D)}_{N 1}&\cdots
&\eta^{(D1)}_{N 1}&\cdots& \eta^{(DD)}_{N 1}
\end{array}
\right),\\\label{matr_A}
\hat A^{(i)}&=&\left(
\begin{array}{cccccccccc}
1 &\cdots &\hat a^{(i;1D)}_{1 }&\hat a^{(i;21)}_{1 }&\cdots& \hat a^{(i;2D)}_{1 }&\cdots
&\hat a^{(i;D1)}_{1 }&\cdots& \hat a^{(i;DD)}_{1 }\cr
1 &\cdots &\hat a^{(i;1D)}_{2 }&\hat a^{(i;21)}_{2 }&\cdots& \hat a^{(i;2D)}_{2 }&\cdots
&\hat a^{(i;D1)}_{2 }&\cdots& \hat a^{(i;DD)}_{2 }\cr
\cdots&\cdots&\cdots&\cdots&\cdots&\cdots&\cdots&\cdots&\cdots&\cdots\cr
1&\cdots &\hat a^{(i;1D)}_{N }&\hat a^{(i;21)}_{N }&\cdots& \hat a^{(i;2D)}_{N }&\cdots
&\hat a^{(i;D1)}_{N }&\cdots& \hat a^{(i;DD)}_{N }
\end{array}
\right)
\end{eqnarray}
(for the sake of brevity, we do not split matrices $\hat a^{(j;m_1m_2)}$ into 
$\hat a^{(j;m_11)}$ and $\hat a^{(j;1m_2)}$).
 The elements of these matrices read:
\begin{eqnarray}\label{matr_xi}
&&
\hat\xi_{\beta j}|_{j=(m_2-1) D + m_1}  =\eta^{(m_1m_2)}_{\beta 1},\\\nonumber
&&
\hat A^{(i)}_{\beta j}|_{j=(m_2-1) D + m_1}  =\hat a^{(i;m_1m_2)}_{\beta }=
\hat a^{(i;m_11)}_{\beta }\hat a^{(i;1m_2)}_{\beta },
\\\nonumber
&&
m_1,m_2=1,\dots,D,
\;\;\beta=1,\dots,N,\;\;
j=1,\dots, \tilde D.
\end{eqnarray}
Finally,  we can write system (\ref{hata}) in the following matrix form: 
\begin{eqnarray}\label{fa}
 a^{(i)} \hat \xi =\hat A^{(i)} .
\end{eqnarray}
Eq.(\ref{fa}) may be uniquely solved for    $a^{(i)}$ if 
\begin{eqnarray}\label{ND}
N= D^2, \;\;\det \hat \xi \neq 0.
\end{eqnarray}
In this case we have:
\begin{eqnarray}\label{fa_sol}
 a^{(i)} =\hat A^{(i)} \hat \xi^{-1}.
\end{eqnarray}
Herewith, the elements $\hat a^{(i;1m_2)}_{\beta}$ and $\hat a^{(i;m_11)}_{\beta}$  are arbitrary parameters. 
If $N>D^2$, then matrices $a^{(i)}$ are not unique and some elements of the matrices  $a^{(i)}$ 
can be arbitrary as well. Having 
determined 
matrices $a^{(i)}$, 
we find matrices $\hat a^{(i)}$ through eq.(\ref{hatai}) and thus 
fix   all coefficients in  systems (\ref{I_Z2gen21f}) and (\ref{tZ2gen2}) in terms of 
$\hat a^{(i;1m_2)}_{\beta}$ and $\hat a^{(i;m_11)}_{\beta}$. Consequently, 
matrices 
$L^{(m_1)}$ and $R^{(m_2)}$, $m_1,m_2=2,\dots,D$, constructed as solutions of 
(\ref{I_Z2gen21f}) and (\ref{tZ2gen2}), depend on 
$\hat a^{(i;1m_2)}_{\beta}$ and $\hat a^{(i;m_11)}_{\beta}$ as well.

\subsection{Eq.(\ref{S_Q_simple_d_nl_U_h_matr}) as a reduction of eq.(\ref{S_Q_simple}): 
solution of eq.(\ref{I_hata}) with $N<D^2$}
\label{Section:DlessN}

Since $N\ge  D^2\sim K^4$ in Sec.\ref{Section:redZ}, the matrix dimensionality $N$ is very large and 
increases very fast with an increase in the  number $K$ of  independent functions $\hat g^{(k)}(\lambda)$. 
In this section, we decrease the matrix 
dimensionality and reduce  nonlinear PDE (\ref{S_Q_simple}) to eq.(\ref{S_Q_simple_d_nl_U_h_matr}).

 Let $N< D^2$.
Now  $\hat \xi$ and  $\hat A^{(i)}$ are rectangular $N\times  D^2$ 
matrices and they can  be represented in the 
  following block forms:
\begin{eqnarray}\label{xi12}
\hat \xi =\Big(
\hat\xi_1 \;\; \hat \xi_2 \Big), \;\; \hat A^{(i)} =
\Big(\hat A^{(i)}_1 \;\; \hat  A^{(i)}_2 \Big),
\end{eqnarray}
where $\hat\xi_1$ and $\hat A^{(i)}_1$ are the square $N\times N$ matrices, while 
$\hat\xi_2$ and $\hat A^{(i)}_2$ are the rectangular $ N\times (D^2-N)$ matrices.
We split eq.(\ref{fa})  into two following equations:
\begin{eqnarray}\label{fa2}
&&a^{(i)}\hat \xi_1 =\hat A^{(i)}_1 ,\\\label{fa22}
&&a^{(i)}\hat \xi_2 =\hat A^{(i)}_2,\;\;i=1,\dots,K.
\end{eqnarray}
Let $\hat \xi_1$  be invertible. Without the loss of generality, we take $\hat \xi_1=I_N$.
Then  eq.(\ref{fa2}) defines $ a^{(i)}$:
\begin{eqnarray}\label{fa21}
 a^{(i)}= \hat A^{(i)}_1 ,\;\;i=1,\dots,K.
\end{eqnarray}
In particular, if $N=D$, then $a^{(i)}_{\alpha m_1} = a^{(i;1m_1)}_\alpha$.
 Substituting eq.(\ref{fa21})  into eq.(\ref{fa22}), we obtain:
\begin{eqnarray}\label{fa222}
\hat A^{(i)}_1\hat \xi_2=\hat A^{(i)}_2,\;\;i=1,\dots,K.
\end{eqnarray}
Eq.(\ref{fa222}) with $i=1$ defines $\hat \xi_2$ (we assume that  
$\det \,\hat A^{(i)}_1 \neq 0$):
\begin{eqnarray}\label{fa22xi}
\hat \xi_2=(\hat A^{(1)}_1)^{-1} \hat A^{(1)}_2.
\end{eqnarray}
It is obvious, that eqs. (\ref{fa222}) with $i>1$ hold if 
\begin{eqnarray}\label{AAA}
\hat A^{(i)}_2 = \Pi^{(i)} \hat A^{(1)}_2,\;\;\hat A^{(i)}_1 = \Pi^{(i)} \hat A^{(1)}_1,\;\;\Pi^{(1)}=I_N,
\end{eqnarray}
where $\Pi^{(i)}$ are constant matrices and $I_N$ is the $N\times N$ identity matrix. 
However, one has to remember
that $ \hat A^{(i)}_1$ and $ \hat A^{(i)}_2$ are blocks of the same matrix $\hat A^{(i)}$ 
with elements having a certain structure. 
 Therefore, not any constant matrices $\Pi^{(i)}$ can be taken in relations (\ref{AAA}). 
 The allowed matrices $\Pi^{(i)}$ are those  that, {{} after multiplying some matrix  from the left, 
 just rearrange rows of this matrix.} 
 Thus, all matrices $\hat A^{(i)}$, $i=1,\dots,K$,  consist of rows of matrix
 $\hat A^{(1)}$ taken in different orders (the same holds for the matrices
 $\hat A^{(i)}_k$, $k=1,2$).  

\paragraph{Analysis of eqs.(\ref{I_Z2gen21f}) and (\ref{tZ2gen2}).}

Having relations (\ref{AAA}) among $\hat A^{(i)}_k$, $k=1,2$, and, consequently, the similar relations  among
$a^{(i)}$, $\hat a^{(i)}$ and $\hat a^{(i;m_1m_2)}$,
\begin{eqnarray}\label{AAA20}
a^{(i)}=\Pi^{(i)} a^{(1)},\;\;\;\hat a^{(i)}= \hat a^{(1)}(\Pi^{(i)})^T ,\;\;
\hat a^{(i;m_1m_2)}=\Pi^{(i)} \hat a^{(1;m_1m_2)}(\Pi^{(i)})^T,
\end{eqnarray}
 we establish the equivalence between equations (\ref{I_Z2gen21f}) with  $i=j$, i.e.,
 only one of them is independent:
 \begin{eqnarray}\label{Z1gen2_Pi}
 &&
 \sum_{m_1=1}^{D} L^{(m_1)}
 \hat a^{(i)}  \hat a^{(i;m_11)}=0\;\; \Leftrightarrow \sum_{m_1=1}^{D} L^{(m_1)}
 \hat a^{(1)} (\Pi^{(i)})^T \Pi^{(i)} \hat a^{(1;m_11)}=0 \;\;\Rightarrow \\\nonumber
 &&\sum_{m_1=1}^{D} L^{(m_1)}
 \hat a^{(1)}  \hat a^{(1;m_11)}=0 .
 \end{eqnarray}
 Eqs.(\ref{I_Z2gen21f}) with $i\neq j$  read
 \begin{eqnarray}\label{Z2gen21_Pi}
 \sum_{m_1=1}^{D}L^{(m_1)}
 \hat a^{(1)} (\Pi^{(i)})^T \Pi^{(j)} \hat  a^{(1;m_11)}=0,\;\;i\neq j,\;\;i,j=1,\dots,K.
 \end{eqnarray}
 Similarly,  eqs. (\ref{tZ2gen2})  get the following form
\begin{eqnarray}
\label{tZ2gen2_Pi}
 &&\sum_{m_2=1}^{D} 
   \hat a^{(1;1m_2)}(\hat a^{(1)})^TR^{(m_2)}=0,\\
 \label{tZ1gen21_Pi}
 &&
\sum_{m_2=1}^{D}
 \hat  a^{(1;1m_2)}  (\Pi^{(j)})^T \Pi^{(i)} (\hat a^{(1)})^T  R^{(m_2)}=0,\;\;i\neq j,\;\;i,j=1,\dots,K.
\end{eqnarray}
Notice that not all equations in  systems (\ref{Z2gen21_Pi})  and (\ref{tZ1gen21_Pi}) are independent. 
We consider their dependency  using a particular representation of $\Pi^{(i)}$ in 
terms of the  matrix $\Pi$ (\ref{Pi}) which shifts the rows:
  $\Pi^{(i)} = \Pi^{i-1}$, $i\ge 1$.
Then eqs.(\ref{AAA20}) read
\begin{eqnarray}\label{AAA2}
a^{(i)}=\Pi^{i-1} a^{(1)},\;\;\;\hat a^{(i)}=\hat a^{(1)}(\Pi^T)^{i-1},\;\;\hat a^{(i;m_1m_2)}=\Pi^{i-1} \hat a^{(1;m_1m_2)}(\Pi^T)^{i-1}.
\end{eqnarray}
 Only those of eqs.(\ref{Z2gen21_Pi}) (and (\ref{tZ1gen21_Pi})) are independent which have different 
values $i-j$ (this increment  can be either positive or negative). 
Thus, we reduce the system (\ref{Z1gen2_Pi},\ref{Z2gen21_Pi}) to
\begin{eqnarray}
\label{Z2gen_red}
&&\sum_{m_1=1}^{D}L^{(m_1)}
 \hat a^{(1)} \hat  a^{(i;m_11)}=0\;\;\Rightarrow \;\;
\sum_{m_1=1}^{D}\tilde L^{(m_1)}
  \Pi^{i-1} \hat  a^{(1;m_11)}=0,\;\;i=1,\dots, K, \\\nonumber
&&\sum_{m_1=1}^{D}L^{(m_1)}
 \hat a^{(i)} \hat  a^{(1;m_11)}=0\;\;\Rightarrow \;\;
\sum_{m_1=1}^{D}\tilde L^{(m_1)}
 (\Pi^T)^{i-1}  \hat  a^{(1;m_11)}=0,\;\;i=2,\dots, K ,\\\nonumber
&&
\tilde L^{(m_1)} = L^{(m_1)}  \hat a^{(1)},
\end{eqnarray}
which is a system of $2 K-1$ equations.
Alternatively, this system  can be
represent   by the single formula
\begin{eqnarray}\label{Z2gen_red_comp}
&&
\sum_{m_1=1}^{D}\tilde L^{(m_1)}_{\alpha\beta}
 \hat  a^{(1;m_11)}_{\beta\pm  i} =0\;\;\;\Rightarrow \;\;\; 
 \sum_{m_1=1}^{D}\tilde L^{(m_1)}_{\alpha(\beta\pm  i)}
 \hat  a^{(1;m_1 1)}_{\beta} =0
 , \\\nonumber
 &&
 \hat  a^{(1;m_11)}_{N+i} =\hat  a^{(1;m_11)}_i,\;\;
 \hat  a^{(1;m_11)}_{-i} =\hat  a^{(1;m_11)}_{N-i},\\\nonumber
 &&
 \tilde L^{(m_1)}_{\alpha (N+i)} =\tilde L^{(m_1)}_{\alpha i},\;\;
\tilde L^{(m_1)}_{\alpha(-i)} =\tilde L^{(m_1)}_{\alpha(N-i)},\;\;
i=0,1,\dots, K-1.
\end{eqnarray}
Thus, the total number of equations is reduced from  $K^2$ in  system (\ref{I_Z2gen21f}) to $(2 K -1)$ 
in system (\ref{Z2gen_red_comp}). 
Consequently, we should take $D=  2 K $
to  provide the solvability of system (\ref{Z2gen_red_comp}) with respect to $\tilde L^{(m_1)}$.

Similarly,  the system 
(\ref{tZ2gen2_Pi}-\ref{tZ1gen21_Pi}) reduces to
\begin{eqnarray}
\label{tZ2gen_red}
&&\sum_{m_2=1}^{D}
 \hat  a^{(i;1m_2)}(\hat a^{(1)})^T R^{(m_2)}=0\;\;\Rightarrow \;\;
\sum_{m_2=1}^{D}
 \hat  a^{(1;1m_2)} (\Pi^T)^{i-1} \tilde R^{(m_2)}=0,\;\;i=1,\dots, K, \\\nonumber
&&\sum_{m_2=1}^{D}
  \hat  a^{(1;1m_2)} (\hat a^{(1)})^T R^{(m_2)}=0\;\;\Rightarrow \;\;
\sum_{m_2=1}^{D}
  \hat  a^{(1;1m_2)}\Pi^{i-1}\tilde R^{(m_2)}=0,\;\;i=2,\dots, K ,\\\nonumber
  &&
  \tilde R^{(m_1)} =(\hat a^{(1)})^T R^{(m_1)} 
\end{eqnarray}
which is also a system of $2 K-1$ equations. Alternatively, this system can be written as
\begin{eqnarray}\label{tZ2gen_red_comp}
&&
\sum_{m_2=1}^{D}
 \hat  a^{(1;1m_2)}_{\alpha\pm  i}\tilde R^{(m_2)}_{\alpha\beta} =0\;\;\;\Rightarrow \;\;\;
 \sum_{m_2=1}^{D}
 \hat  a^{(1;1m_2)}_{\alpha}\tilde R^{(m_2)}_{(\alpha\pm  i)\beta} =0,
 \\\nonumber
 &&
 \hat  a^{(1;1 m_2)}_{N+i} =\hat  a^{(1;1m_2)}_i,\;\;
 \hat  a^{(1;1m_2)}_{-i} =\hat  a^{(1;1m_2)}_{N-i}, \\\nonumber
 &&
 \tilde R^{(m_2)}_{ (N+i)\alpha} =\tilde R^{(m_2)}_{i\alpha },\;\;
\tilde R^{(m_2}_{(-i)\alpha} =\tilde R^{(m_2)}_{(N-i)\alpha},\;\;
i=0,1,\dots, K-1.
\end{eqnarray}
Thus, the total number of equations is reduced from  $K^2$ in
system (\ref{I_tZ1gen21}) to $2 K -1$ in  system (\ref{tZ2gen_red_comp}), and the choice
 $D=  2 K $ provides solvability of system (\ref{tZ2gen_red_comp})
with respect to $\tilde R^{(m_2)}$.

Next, it is important to note that systems  (\ref{Z2gen_red_comp})  and (\ref{tZ2gen_red_comp})
allow us to consider the case of 
diagonal matrices $\tilde L^{(m_1)}$ and $\tilde R^{(m_1)}$, i.e., these matrices can be 
represented  as follows:
\begin{eqnarray}\label{LR_diag}
&&
\tilde L^{(m_1)}_{\alpha\beta}= \delta_{\alpha\beta}\hat  L^{(m_1)}_{\beta},\\\label{LR_diag2}
&&
\tilde R^{(m_2)}_{\alpha\beta}= \delta_{\alpha\beta}\hat  R^{(m_2)}_{\alpha}.
\end{eqnarray}
Now   systems (\ref{Z2gen_red_comp})  and (\ref{tZ2gen_red_comp}) may be
written as, respectively,  eqs.(\ref{Z2gen_red_comp2}) and (\ref{tZ2gen_red_comp2})
which can be
solved for $\hat a^{(1;m_1 1)}$ 
and $\hat a^{(1;1m_2)}$, $m_1,m_2=2,\dots,D$.

Finally, we transform expression (\ref{I_mT})  for $T^{(m_1m_2)}$  to form (\ref{T_red}).  
In virtue of decomposition (\ref{I_aa_gen}),  formula  (\ref{I_mT}) reads
\begin{eqnarray}\label{I_mT2}\label{mT}
T^{(m_1m_2)}(\lambda) =\sum_{j=1}^K 
\tilde g^{(j)}(\lambda) \hat a^{(j;m_11)}\hat a^{(j;1m_2)}.
\end{eqnarray}
Substituting eqs.(\ref{AAA2}) into (\ref{mT}), we result in eq.(\ref{T_red}).

\paragraph{Derivation of eq.(\ref{S_Q_simple_d_nl_U_h_matr}).}
Let us turn to  nonlinear equation (\ref{S_Q_simple}) and write it in the form
\begin{eqnarray}\label{S_Q_simple_d}
\sum_{m_1,m_2=1}^{D} \hat L^{(m_1)}\left( \tilde V_{t_{m_1m_2}} + \tilde V (\hat a^{(1)})^T \xi (\eta^{(m_1m_2)})^T  \hat a^{(1)} \tilde V -
\tilde V (\hat a^{(1)})^T \eta^{(m_1m_2)} \xi^T \hat a^{(1)}\tilde V
\right)\hat R^{(m_2)} =0,\;\;\;
\end{eqnarray}
where 
\begin{eqnarray}\label{I_tV0}
&&
\tilde V=(\hat a^{(1)})^{-1} V((\hat a^{(1)})^T)^{-1}.
\end{eqnarray}
Eq.(\ref{I_tV0}) can be transformed to eq.(\ref{tV}) 
 using eq.(\ref{Vp}) for $V$, eq.(\ref{g}) for $r$, eq.(\ref{hatai}) for $ a^{(i)}$, and eq.(\ref{AAA2}) for 
 $\hat a^{(i)}$.
Taking into account relations (\ref{I_hata}), (\ref{Z2gen_red_comp2}) and (\ref{tZ2gen_red_comp2}) we may rewrite  eq.(\ref{S_Q_simple_d}) in
 components eliminating some of the nonlinear terms:
\begin{eqnarray}\label{S_Q_simple_d_nl}
&&
\sum_{m_1,m_2=1}^{D} \hat L^{(m_1)}_\alpha\left( (\tilde V_{\alpha\beta})_{t_{m_1m_2}} +
\frac{1}{1+Q}\sum_{{{\gamma,\delta=1}\atop{\delta\notin Z(\alpha,\beta,K)}}\atop{\gamma\neq \delta}}^N
s_\gamma \hat a^{(1;m_1m_2)}_\delta \Big(\tilde V_{\alpha\gamma} \tilde V_{\delta\beta} 
-
\tilde V_{\alpha\delta}
 \tilde V_{\gamma\beta} \Big)
\right)\hat R^{(m_2)}_\beta =0,\\\nonumber
&&
\alpha,\beta=1,\dots,N,
\end{eqnarray}
where we introduce the diagonal matrix  $s$ with the diagonal elements
$s_\alpha$,
\begin{eqnarray}\label{s}
s_\alpha=\sum_{\gamma=1}^N a^{(1)}_{\alpha\gamma} \xi_{\gamma 1},
\end{eqnarray}
and use the set of indices $Z(\alpha,\beta,K)$  defined in eq.(\ref{Zlist}).
Assuming the invertibility of the matrix  $a^{(1)}$, we can use eq.(\ref{s}) as the definition of the $N\times1$ matrix   $\xi$ in terms of the arbitrary parameters $s_\alpha$:
\begin{eqnarray}
\label{xi}
\xi = (a^{(1)})^{-1}s p.
\end{eqnarray}
Finally, let us introduce the function $U(t)$ with elements 
\begin{eqnarray}
\label{I_tVU}
U_{\alpha\beta}(t)=\frac{s_\alpha \tilde V_{\alpha\beta}(t)s_\beta}{1+Q} .
\end{eqnarray}
Then eq.(\ref{S_Q_simple_d_nl}) results in the following one:
\begin{eqnarray}\label{S_Q_simple_d_nl_U}
&&
\sum_{m_1,m_2=1}^{D} \hat L^{(m_1)}_\alpha\left( ( U_{\alpha\beta})_{t_{m_1m_2}} +
\sum_{{{\gamma,\delta=1}\atop{\delta\notin Z(\alpha,\beta,K)}}\atop{\gamma\neq \delta}}^N\frac{\hat a^{(1;m_1m_2)}_\delta}{s_\delta}  \Big(
U_{\alpha\gamma} U_{\delta\beta} 
-
 U_{\alpha\delta}
  U_{\gamma\beta} \Big)
\right)\hat R^{(m_2)}_\beta =0,\\\nonumber
&&
\alpha,\beta=1,\dots,N.
\end{eqnarray}
Introducing $N\times 1$ matrix $p=(\underbrace{1\;\dots\;1}_N)^T$ and using decomposition (\ref{I_aa_gen}), 
we may write eq.(\ref{S_Q_simple_d_nl_U}) in  matrix form (\ref{S_Q_simple_d_nl_U_h_matr}).
Finally, using eqs.(\ref{AAA2}) for $\hat a^{(i)}$  and (\ref{s}) for $s_\alpha$, we transform 
eq.(\ref{htg}) for $\tilde g^{(1)}$
into eq.(\ref{tgr}). 

{{}
\section{Appendix B. Richness of solution space}
\label{Section:rich}
\label{Section:appendixB}
 Now we 
discuss the richness of solution space of nonlinear PDEs 
(\ref{S_Q_simple}) and (\ref{S_Q_simple_d_nl_U_h_matr}). 
Since the $t$-dependence is introduced through  the diagonal matrix  
$\varepsilon(\lambda,t)$ given in eq.(\ref{I_chi1}),
the richness of solution space is defined by the  function $\Psi_0$ (\ref{Psi00}):
\begin{eqnarray}\label{Exp}
(\Psi_0(\lambda,\mu;t))_{\gamma\delta} =
{\cal{C}}_{\gamma\delta}(\lambda,\mu)
\exp\left[\sum_{m_2=1}^{D}\sum_{m_1=1}^{D} t_{m_1m_2} \Big(T^{(m_1m_2)}_\gamma (\lambda) -
 T^{(m_1m_2)}_\delta (\mu)\Big)\right].
\end{eqnarray}
Since the number of  arbitrarily introduced variables
 $t_{m_1m_2}$ coincides with the number of  independent combinations of  the 
 functions of spectral parameters in the exponent of expression (\ref{Exp}),
 we have to define the number of such combinations.
 Therewith, the
functions $T^{(m_1m_2)}$ are defined by eq.(\ref{I_mT2}).
}

 \subsection{$s$-dimensionality of eq.(\ref{S_Q_simple})}
  \label{Section:tNs1}
 
 Using expressions (\ref{mT}) for $T^{(m_1m_2)}$,
  we write the argument in the exponent of eq.(\ref{Exp}) as 
\begin{eqnarray}\label{exp}
&&
\sum_{m_1,m_2=1}^{D} t_{m_1m_2} (T^{(m_1m_2)}_\gamma (\lambda) -
 T^{(m_1m_2)}_\delta (\mu)) = \\\nonumber
&&\sum_{m_1m_2=1}^{D}
\sum_{j=1}^K \left(\tilde g^{(j)}_\gamma (\lambda)  \hat a^{(j;m_11)}_\gamma \hat a^{(j;1m_2)}_\gamma t_{m_1m_2}  - \tilde g^{(j)}_\delta (\mu) 
\hat a^{(j;m_11)}_\delta\hat a^{(j;1m_2)}_\delta  t_{m_1m_2}  \right).
\end{eqnarray} 
Remember, that  only $K-1$ functions in  the list $\tilde g^{(i)}$, $i=1,\dots,K$, 
are independent functions of spectral parameters. Thus, in the list  of  functions
$\tilde g^{(i)}_\gamma$,
$\tilde g^{(j)}_\delta$ (with fixed $\gamma$ and $\delta$, $\gamma\neq \delta$, and $i,j=1,\dots,K$)
there are $2(K-1)$ 
independent functions.
  These   $2 (K-1)$  functions introduce   $2 (K-1)$ independent variables $t_{m_1m_2}$, i.e.,
  $s$-dim.$=2(K-1)$.
  If $\gamma=\delta$, then expression (\ref{exp}) involves only $K-1$ 
independent combinations
$\tilde g^{(j)}_\gamma(\lambda) -\tilde g^{(j)}_\gamma(\mu)$, $j=2,\dots,K$, which  introduce 
$K-1$ independent variables $t_{m_1m_2}$.
  Thus,  expression (\ref{Exp})  introduces  up to
  $N(N-1)$ arbitrary functions of $2 (K-1)$ variables and up to $N$ arbitrary functions of 
  $(K-1)$ variables.

  {{}
 \subsection{$s$-dimensionality of eq.(\ref{S_Q_simple_d_nl_U_h_matr})}
  \label{Section:tNs2}
  In the case of eq.(\ref{S_Q_simple_d_nl_U_h_matr}), we shall 
  use expression (\ref{T_red}) for $T^{(m_1m_2)}$ so that the  exponent in eq. (\ref{Exp})
reads
\begin{eqnarray}\label{exp2}
&&
\sum_{m_2=1}^{D}\sum_{m_1=1}^{D} t_{m_1m_2} (T^{(m_1m_2)}_\gamma (\lambda) -
 T^{(m_1m_2)}_\delta (\mu)) = \\\nonumber
&&\sum_{m_2=1}^{D}\sum_{m_1=1}^{D}
\sum_{j=1}^K \left(\tilde g^{(j)}_\gamma (\lambda) 
\Big(\Pi^{j-1}\hat a^{(1;m_11)}\hat a^{(1;1m_2)}(\Pi^T)^{j-1}\Big)_\gamma t_{m_1m_2}  -\right. \\\nonumber
&&\left.
\tilde g^{(j)}_\delta (\mu) \Big(\Pi^{j-1} \hat a^{(1;m_11)}\hat a^{(1;1m_2)}(\Pi^T)^{j-1}\Big)_\delta  t_{m_1m_2}  \right)=\\\nonumber
&&
\sum_{m_2=1}^{D}\sum_{m_1=1}^{D}
\sum_{j=1}^K \left(\tilde g^{(j)}_\gamma (\lambda)  
\hat a^{(1;m_11)}_{\gamma+j-1}\hat a^{(1;1m_2)}_{\gamma+j-1} t_{m_1m_2}  - 
\tilde g^{(j)}_\delta (\mu) \hat a^{(1;m_11)}_{\delta+j-1} \hat a^{(1;1m_2)}_{\delta+j-1}  t_{m_1m_2}  \right).
\end{eqnarray} 
In this case the $s$-dimensionality remains the same (i.e., it equals $2(K-1)$), but the number of arbitrary 
scalar functions of  $2(K-1)$ variables is reduced.

In fact, suppose that $\gamma-\delta=\Delta>0$ (the case with negative $\Delta$ can be treated similarly). 
As was argued above, there are only $K-1$ independent diagonal functions of spectral parameters 
in the list $\tilde g^{(i)}(\lambda)$, $i=1,\dots,K$.
We take  first $K-1$ of them without the loss of generality. Thus, to consider only 
independent functions of spectral parameters in
expression (\ref{exp2}), we cut the sum over $j$ reducing 
the upper  limit from $K$ to $K-1$. This cut exponent 
(after  introducing the parameter $\Delta$) reads:
\begin{eqnarray}\label{exp_red}
&&
\sum_{m_2=1}^{D}\sum_{m_1=1}^{D}
\sum_{j=1}^{K-1} \left(\tilde g^{(j)}_{\delta+\Delta} (\lambda)  
\hat a^{(1;m_11)}_{\delta+\Delta+j-1} \hat a^{(1;1m_2)}_{\delta+\Delta+j-1}t_{m_1m_2}  - 
\tilde g^{(j)}_\delta (\mu) \hat a^{(1;m_11)}_{\delta+j-1} \hat a^{(1;1m_2)}_{\delta+j-1} 
t_{m_1m_2}  \right)=\\\nonumber
&&
\sum_{m_2=1}^{D}\sum_{m_1=1}^{D}\sum_{j=1}^{K-1-\Delta} 
(\tilde g^{(j)}_{\delta+\Delta} (\lambda) -\tilde g^{(j+\Delta)}_{\delta} (\mu)) 
\hat a^{(1;m_11)}_{\delta+\Delta+j-1} \hat a^{(1;1m_2)}_{\delta+\Delta+j-1}t_{m_1m_2} 
+ 
\\\nonumber
&&
\sum_{m_2=1}^{D}\sum_{m_1=1}^{D}\sum_{j=K-\Delta}^{K-1} \tilde g^{(j)}_{\delta+\Delta} (\lambda) 
\hat a^{(1;m_11)}_{\delta+\Delta+j-1}\hat a^{(1;1m_2)}_{\delta+\Delta+j-1}t_{m_1m_2}
-
\\\nonumber
&&
\sum_{m_2=1}^{D}\sum_{m_1=1}^{D}\sum_{j=1}^{\Delta}\tilde g^{(j)}_\delta(\mu) \hat a^{(1;m_11)}_{\delta+j-1}\hat a^{(1;1m_2)}_{\delta+j-1}
t_{m_1m_2}.
\end{eqnarray}
We see that if $\Delta=0$, then there are $K-1$ independent combinations of functions 
of spectral parameters: $(\tilde g^{(j)}_{\delta} (\lambda) -\tilde g^{(j)}_{\delta} (\mu))$
($j=1,\dots,K-1$). If $1\le \Delta \le (K-2)$, then the number of independent combinations becomes 
 $K-1 +\Delta$:  
 $(\tilde g^{(j)}_{\delta+\Delta} (\lambda) -\tilde g^{(j+\Delta)}_{\delta} (\mu))$ ($j=1,\dots,K-1-\Delta$),   
$\tilde g^{(j)}_{\delta+\Delta} (\lambda)$ ($j= K-\Delta, \dots, K-1$),
$\tilde g^{(j)}_\delta(\mu)$, $j=1,\dots,\Delta$.
If $\Delta \ge (K-1)$, then the number of independent functions of spectral parameter reaches its maximal value 
$2(K-1)$.

All in all, expression (\ref{Exp}) introduces up to 
 $N$ functions 
 of $K-1$ variables, up to $2(N-\Delta)$ functions of 
 $K-1+\Delta$ variables ($1 \le \Delta\le K-2$), and up to 
 $N^2-N-2\sum_{\Delta=1}^{K-2} (N-\Delta) =(N-K+2) ( N - K +1) $ functions of $2(K-1)$ variables. 
}

 {{}
\subsection{Number of arbitrary functions in  solution space  under Hermision reduction}
The Hermitian reduction reduces the number of arbitrary functions in the solution 
space. Owing to the relation 
(\ref{V02}), the number of arbitrary complex  functions of $2(K-1)$ 
variables reduces by two times in solution spaces of  both  eq.(\ref{S_Q_simple}) under the Hermitian reduction 
and 
eq.(\ref{S_Q_simple_d_nl_U_h}) in comparison with the solution spaces of, respectively,
 eq.(\ref{S_Q_simple}) and 
eq.(\ref{S_Q_simple_d_nl_U_h_matr}). 
Thus, regarding the  eq.(\ref{S_Q_simple}) under the Hermitian reduction, 
this number is  $N(N-1)/2$ instead of $N(N-1)$. 
The number of arbitrary functions of $(K-1)$ variables   remains $N$ in  both equations. 
The number of other functions in solution space of eq.(\ref{S_Q_simple_d_nl_U_h}) is reduced by two times. 
}

\section{Appendix C. Relations among $t$-, $N$- and $s$-dimensionalities of 
eqs.(\ref{S_Q_simple}) and (\ref{S_Q_simple_d_nl_U_h_matr})}
\label{Section:DKN}
\label{Section:appendixC}
Here   we collect 
  results regarding the relations among such important parameters of 
  eqs.(\ref{S_Q_simple}) and (\ref{S_Q_simple_d_nl_U_h_matr})  as $t$-, $N$- and $s$-dimensionalities. 
  We show    
  that all parameters can be expressed in terms of the parameter $K$ 
  (the number of independent diagonal matrix 
functions $g^{(i)}(\lambda)$  in the  definition of the function $r(\lambda)$, see eq.(\ref{g})). 

\paragraph{Relations among parameters in eq.(\ref{S_Q_simple}).} In Sec.\ref{Section:tNs1} we found that 
 $s$-dim.$=2(K-1)$ for eq.(\ref{S_Q_simple}).

 Next, the   matrices $L^{(m_1)}$ (and $R^{(m_1)}$), $m_1=1,\dots,D$, can be  considered 
as solutions  to the system of $K^2$ matrix equations  (\ref{I_Z2gen21f}) (and   (\ref{I_tZ2gen22}))
if the number of equations in system (\ref{I_Z2gen21f}) (and   (\ref{I_tZ2gen22}))
does not exceed the number of matrices  $L^{(m_1)}$ 
(and $R^{(m_1)}$), $m_1=1,\dots,D$, i.e., we take
\begin{eqnarray}\label{DKgen}
D = K^2 +1.
\end{eqnarray}
Thus, the $t$-dim.$=D^2= (K^2+1)^2$. 

Finally,  it was shown in Sec.\ref{Section:redZ} that
 system of  diagonal equations (\ref{hata}) is solvable for  
 $a^{(i)}$ if $N \ge D^2$, i.e.,
 $N$-dim.$\ge (K^2+1)^2$. 

Collecting the above results, we write the following  relations among three dimensionalities:
\begin{eqnarray}
{\mbox{$t$-dim.}} \sim  ({\mbox{$s$-dim.}})^4,\;\;{\mbox{$N$-dim.}} \gtrsim  ({\mbox{$s$-dim.}})^4.
\end{eqnarray}

\paragraph{Relations among parameters in eq.(\ref{S_Q_simple_d_nl_U_h_matr}).} In Sec.\ref{Section:tNs2}, 
we found that  $s$-dim.$=2(K-1)$. 
 
Next, in Sec.\ref{Section:DlessN}, we  take $D=2 K$ 
to provide the solvability of equations (\ref{Z2gen_red_comp2}) and (\ref{tZ2gen_red_comp2}) with respect to, 
respectively,  
$\hat L^{m_1}$ and $\hat R^{m_1}$, $m_1=2,\dots,D$. Thus,
$t$-dim.$=D^2 =4 K^2$.

The $N$-dimensionality of  PDE (\ref{S_Q_simple_d_nl_U_h_matr})
defines the structure of the nonlinear terms in eq. (\ref{S_Q_simple_d_nl_U}).
To describe this effect, we turn to eq. (\ref{S_Q_simple_d_nl_U}).
It can be readily checked, that the nonlinear part disappears from eq.(\ref{S_Q_simple_d_nl_U}) if $N <2 K$. 
If $N=2K$, then the nonlinear part exists only if $\alpha=\beta$, i.e., only 
the diagonal elements $U_{\alpha\alpha}$ satisfy the nonlinear PDE. In general, 
if $N> 2K$, then the nonlinear part exists if $\alpha=\sigma \pm i$, $\beta = \sigma \pm j$ 
with $i, j =0,1\dots,N-2 K$.
Thus, if $2K\le N < 4 K-1$, we have a ''partial nonlinearity'',  i.e some elements of
the matrix $U$ satisfy the linear equations. Note that, in many cases, this is a 
"hidden linearity", i.e., all nonlinear terms (which are quadratic by construction)
involve those elements of the matrix $U$ that  satisfy the linear PDEs. 
The system becomes completely nonlinear if $2 (N-2 K)+1 \ge N$, i.e., $N\ge 4 K-1$.
}

Thus, for the completely nonlinear matrix PDE, 
\begin{eqnarray}
{\mbox{$t$-dim.}} \sim  ({\mbox{$s$-dim.}})^2,\;\;{\mbox{$N$-dim.}} \gtrsim  ({\mbox{$s$-dim.}}).
\end{eqnarray}

\section{Appendix D. Diagonal eq.(\ref{rBr}) with diagonal $C^{(m)}$: classical 
(2+1)-dimensional $n$-wave equation}
\label{Section:appendixD}
According to Sec.\ref{Section:Psi}, eq.(\ref{rBr}) is the principal equation 
in our algorithm. It defined what kind of nonlinear PDEs can be derived. Here 
we consider such solution to this equation 
 that  generates the  classical (2+1)-dimensional $n$-wave equation. 
 
Let eq. (\ref{rBr}) be diagonal. This implies that 
matrices $R(\lambda,\mu)$,  $r(\lambda)$ and $\tilde r(\mu)$ defined by eqs.(\ref{R_r_r}) 
are  diagonal as well.
Let
\begin{eqnarray}
\label{TT_case1}
T^{(m)}(\lambda) =\lambda C^{(m)},\;\;\;\tilde T^{(m)}(\mu) = \mu C^{(m)}
\end{eqnarray}
(here $m=(m_1m_2)$).
Then $C^{(m)}$ may be canceled  from   eq.(\ref{rBr}) which yields 
\begin{eqnarray}\label{R_case1}
R(\lambda,\mu)=\frac{r(\lambda) \tilde r(\mu)}{\mu-\lambda}.
\end{eqnarray}
Since $R$, $r$ and $\tilde r$ are diagonal, 
conditions (\ref{S_constrain0}) and (\ref{S_constrain1}) (or (\ref{T1_2}) and (\ref{T2_2}))  are satisfied if
\begin{eqnarray}
\label{T1_T}
&&
\sum_{m_1}L^{(m_1)} C^{(m_1m_2)}  =0,\\
\label{T1_C2}
&&
\sum_{m_2}
 C^{(m_1m_2)} R^{(m_2)}
=0.
\end{eqnarray}
Owing to eqs.(\ref{TT_case1}), the number of arguments in arbitrary 
functions (parameterizing solution space) is restricted. In fact, 
eq.(\ref{exp}) now reads
\begin{eqnarray}\label{exp_cl}
&&
\sum_{m_1,m_2=1}^{D} t_{m_1m_2} (T^{(m_1m_2)}_\gamma (\lambda) -
 T^{(m_1m_2)}_\delta (\mu)) = 
\sum_{m_1,m_2=1}^{D} t_{m_1m_2} \Big( C^{(m_1m_2)}_\gamma  \lambda-
 C^{(m_1m_2)}_\delta \mu\Big)
 .
\end{eqnarray} 
We see that only two independent spectral  parameters $\lambda$ and $\mu$ appear in this expression  
if $\delta\neq \gamma$. If $\delta=\gamma$, then there is only one spectral parameter $\lambda-\mu$. 
Thus we may introduce $N(N-1)$ arbitrary functions of two independent
variables and $N$ arbitrary functions of one independent variable in the solution space. 
This means that we may provide the full solution space only to 
$(2+1)$ dimensional nonlinear PDE with $N(N-1)$  scalar fields.
To obtain the classical form of this PDE, we take the
real diagonal  matrices $C^{(m_1m_2)}$, $L^{(m_1)}$ and $R^{(m_2)}$ with
\begin{eqnarray}
&&
L^{(m_1)}=R^{(m_1)},\;\;C^{(m_1m_1)}=0,\;\;C^{(m_1m_2)}=-C^{(m_2m_1)},\;\;t_{m_1m_2}=-t_{m_2m_1},\\\nonumber
&&
L^{(1)}=C^{(23)}\equiv C^{(1)}=I_N,\;\;
L^{(2)}=-C^{(13)} \equiv -C^{(2)},\;\;L^{(3)}=C^{(12)}\equiv C^{(3)}
\end{eqnarray}
and introduce the new variables 
\begin{eqnarray}
t^{(1)}\equiv -t^{(23)},\;\;
t^{(2)}\equiv t^{(13)},\;\;t^{(3)}\equiv t^{(12)}.
\end{eqnarray}
Then eq.
(\ref{S_Q}) becomes  classical (2+1)-dimensional nonlinear PDEs (\ref{Nwave}) with
three independent  variables $t_i$, $i=1,2,3$ \cite{Kaup1,Kaup2}.

{{}
\section{Appendix E. Remark on higher order nonlinear PDEs}
\label{Section:appendixE}

As was mentioned in Introduction, Sec.\ref{Section:introduction}, the proposed dressing algorithm can 
be simply generalized to construct 
higher order nonlinear PDEs. For this purpose, we  have to replace the first external constraint  
(\ref{S_constrain0}) 
with another one. 
To explain the way of doing  it, we recall that external constraint  (\ref{S_constrain0}) is introduced
to construct the 
linear integral homogeneous equation with the kernel $\Psi + {\cal{I}}$ (see eq.(\ref{S_lin_2})) 
combining  the first 
derivatives of equation 
(\ref{Psi}). Similarly, we may introduce the first  constraint from the requirement to obtain  
the linear integral homogeneous equation with the kernel $\Psi + {\cal{I}}$ combining   the second, 
third (and so on) 
derivatives of equation 
(\ref{Psi}), thus obtaining the second-, third-order  (and so on) multidimensional matrix nonlinear  PDEs. 
The second external constraint (\ref{S_constrain1}) is introduced to obtain the nonlinear PDE for the 
field $V$ (see eq.(\ref{S_Q})).  In the case of higher order PDE, the modified 
first external constraint forces  the appropriate modification of the  second external constraint. 
More detailed description of higher order nonlinear PDEs and associated dressing algorithm  remains beyond 
the scope of this paper. 
}



\begin{thebibliography}{99}



\bibitem{GGKM}
C.S.Gardner, J.M.Green, M.D.Kruskal, R.M.Miura, Phys.Rev.Lett.
{\bf 19},   1095 (1967)


\bibitem{ZMNP}
V.E.Zakharov, S.V.Manakov, S.P.Novikov and L.P.Pitaevsky, 
{\it Theory of Solitons. The Inverse Problem Method}, Plenum Press (1984)

\bibitem{AC}
M.J.Ablowitz and P.C.Clarkson, {\it Solitons, Nonlinear Evolution Equations and Inverse Scattering}, 
Cambridge University Press, Cambridge, 1991



\bibitem{ZS1}
V.E.Zakharov and A.B.Shabat, Funct.Anal.Appl. {\bf 8}, 43 (1974) 

\bibitem{ZS2}
V.E.Zakharov and A.B.Shabat, Funct.Anal.Appl. {\bf 13},  13  (1979)

\bibitem{ZM}
V.E.Zakharov and S.V.Manakov, Funct.Anal.Appl. {\bf 19},  11 (1985)

\bibitem{BM}
L.V.Bogdanov and S.V.Manakov, J.Phys.A:Math.Gen. {\bf 21},
 L537 (1988)


\bibitem{K}
B. Konopelchenko, {\it Solitons in Multidimensions}, World Scientific, Singapore (1993)


 
 \bibitem{ZS}
A.I.Zenchuk and P.M.Santini, 
J. Phys. A: Math. Gen. {\bf 39},
5825 (2006)
   
\bibitem{Whitham} J. B. Whitham, {\it Linear and Nonlinear Waves}, Wiley, NY, 1974

\bibitem{SZ} P. M. Santini and A. I. Zenchuk,
Phys. Lett.
{\bf A 368}, 48  (2007)

 

\bibitem{T1}
S.P. Tsarev, Sov. Math. Dokl. {\bf 31} No.3,  488 (1985) 

\bibitem{DN}
 B.A.Dubrovin, S.P. Novikov, Russian Math. Surveys {\bf 44} No.6, 35 (1989)
 
 \bibitem{T2}
S.P.Tsarev, Math. USSR Izv. {\bf 37}, 397 (1991)

 \bibitem{F}
 E.V. Ferapontov, Teor. Mat. Fiz. {\bf 99}, 257 (1994)


\bibitem{HopfCole}
Hopf E,  Commun. Pure Appl. Math. {\bf 3} 201 (1950),
Cole J D,  Q. Appl. Math. {\bf 9}, 225 (1951)

\bibitem{Calogero1}
 Calogero F in What is Integrability ed V E Zakharov (Berlin: Springer) (1990) 1

\bibitem{Calogero2}
 Calogero F and Xiaoda Ji, J. Math. Phys. {\bf 32}, 875 (1991)

\bibitem{Calogero3}
 Calogero F and Xiaoda Ji, J. Math. Phys. {\bf 32}, 2703 (1991) 

\bibitem{Calogero4}
Calogero F, J. Math. Phys. {\bf 33}, 1257 (1992)



\bibitem{Santini}
 P.M.Santini, Inverse Problems {\bf 8}, 285 (1992)


\bibitem{W}
R.S.Ward,   Phys.Lett.A {\bf 61}, 81 (1977)

\bibitem{BZ}
A.A.Belavin  and V.E.Zakharov,  Phys.Lett.B {\bf 73}, 53  (1978)


\bibitem{MS1}
S.V.Manakov and P.M.Santini, Phys. Lett. A {\bf 359}, 613 (2006)

\bibitem{MS2}
S. V. Manakov and P. M. Santini, JETP Letters {\bf 83},  462 (2006)
 
\bibitem{KAR}
 B. Konopelchenko, L. Martinez Alonso and O. Ragnisco,  J.Phys. A: Math. Gen. {\bf 34}, 10209 (2001)



 \bibitem{Z3}
 A.I. Zenchuk, J. Phys. A: Math.Gen. {\bf 37},  6557 (2004)

  \bibitem{Z4}
 A.I. Zenchuk, Phys.Lett.A {\bf 375}, 2704 (2011)

\bibitem{Kaup1}
D.J.Kaup,
 Stud. Appl. Math. {\bf 62},  75 (1980)
  
 
\bibitem{Kaup2} 
    D.J.Kaup,
   Physica D {\bf 1}, 45 (1980)
   


\bibitem{Z2010}
A.I.Zenchuk, 
J. Phys. A: Math. Theor. {\bf 43}, 245203 (2010)






\end{thebibliography}
\end{document}